\documentclass[referee]{aa}
\usepackage[dvips]{graphics}
\usepackage{latexsym}

\begin{document}

\thesaurus{3(11.04.1;11.05.1;11.05.2;11.06.2;11.16.1)}

\title{The morphology of extremely red objects}

\author{
	G. Moriondo \inst{1}
	\and
	A. Cimatti \inst{1}
	\and
	E. Daddi
	\inst{2} 
}
\institute{ 
	Osservatorio Astrofisico di Arcetri,
	Largo E. Fermi 5, I-50125 Firenze, Italy
%	\and  
%  	C.N.A.A. 
%	Viale del Parco Mellini 84, I-00136 Roma, Italy
 	\and  
	Universit\`a degli Studi di Firenze, 
	Dipartimento di Astronomia e Scienza dello Spazio,
	Largo E. Fermi 5, I-50125 Firenze, Italy
%	\and 
} 

\offprints{gmorio@arcetri.astro.it}
\date{Received ; accepted }

\maketitle
\markboth{Moriondo et al.: ERO morphology }{Moriondo et al.: ERO morphology}

\begin{abstract}

We present a quantitative study of the morphology of 41 
% high--$z$
Extremely Red Objects (EROs). The analysis is based on deep otical and 
near--infrared images from the Hubble Space Telescope public archive, 
and performed by fitting to each galaxy image
a PSF--convolved bi--dimensional model brightness distribution.
Relying both on the visual inspection of the data and on the results 
of the fitting procedure, we are able to determine the fraction of 
irregular and/or interacting EROs, and to identify those that more
closely resemble local ellipticals. To the former class, whose members are
probably high--redshift dusty starburst, belongs about 15\% of the 
whole sample, whereas the elliptical--like objects are between 50 and 
80\% of the total. A few galaxies, although characterized by a compact
morphology, are best fitted by an exponential distribution, more typical of
local spirals. Our data also suggest that irregular EROs are
found predominantly in the field, and that -- on average --
they tend to be characterized by the reddest colors. Finally, we plot the 
rest--frame Kormendy Relation ($\mu_e$ vs. $R_e$) for a sample of 6 EROs 
with spectroscopic redshifts ($z \sim 1.3$), and estimate its evolution 
with respect to the local relation.

\end{abstract}

\section{Introduction}

Among the variety of objects discovered so far at high--redshift, a 
special class is represented by the so called Extremely Red Objects 
(EROs hereafter),
characterized by moderately faint near--IR magnitudes ($K \sim
18-20$), and extremely red optical--infrared colors (e.g, $R-K > 5$, see
for example Elston et al. \cite{elston}, McCarthy et al. \cite{macca}, 
Hu \& Ridgway \cite{hu}).
The observed colors and luminosities place this class of objects at  
$z \geq 1$, an hypothesis confirmed in a few cases by a direct
spectroscopic measurement of the redshift (Graham \& Dey \cite{grdey},
Spinrad et al. \cite{spin}, Stanford et al \cite{stan} -- S97 hereafter,
Liu et al. \cite{liu}).
A twofold interpretation of such observational properties is possible:
EROs can be either high--redshift starburst galaxies reddened
by a large amount of dust, or passively evolving high--$z$ ellipticals
characterized by old stellar populations ($\geq 1$ Gyr).

The importance of assessing the ERO nature and determining their 
space density is clear: the epoch of formation of massive elliptical
galaxies is a crucial test for the standard hierarchical models for
structure formation (e.g.: White \& Rees \cite{wr}, Kauffmann et 
al. \cite{kauff}), 
which predict such objects to have formed relatively late 
%($z \leq ? $) 
from the merging of smaller--size objects
(presumably disk galaxies). A large density of high redshift evolved 
ellipticals would imply severe revision to the hierarchical theories.
The other relevant question is the global star formation history:
calculations based on the observed rest frame UV flux (e.g. Madau et al.
\cite{madau}, Connolly et al. \cite{conno}) might be significantly 
underestimated, if a large fraction
of the overall star formation at high redshift takes place in highly
obscured starburst galaxies (e.g. Steidel et al. \cite{steid99}; 
Barger et al. \cite{barg00}). 

One way to disentangle this ambiguity is provided, in some cases, by 
near--infrared spectroscopy, in particular if the ERO spectrum exhibits 
features revealing star--formation activity, such as the redshifted 
H$_\alpha$ line; this is the case, for example, for the galaxy
HR10 (Graham \& Dey \cite{grdey}, Dey et al. \cite{dey99}). 
More recently, deep near--infrared spectroscopy allowed to
classify two more galaxies as likely starburst -- although their spectra
lack of spectral features -- from the amount of reddening required 
to explain their overall spectral energy distribution  (Cimatti et al. 
\cite{cimanew}).
A different kind of test is provided by observations in the submm 
waveband, which traces the thermal emission by dust in the starbursts;
this method was successful in the case of HR10 (Cimatti et al. 
\cite{cimatti}, Dey et al. \cite{dey99}) whose detection allowed its
non--ambiguous classification as a dust reddened starburst, a result furtherly 
confirmed by the observation of its CO emission (Andreani et al. 
\cite{andre}).
Other objects, first detected in the submm, have afterwards
turned out to be EROs (Smail et al. \cite{smail}; Gear et al. 
\cite{gear}). 

When images of sufficient spatial resolution are available, however,
the most direct way to distinguish between the two classes is
their morphology: elliptical galaxies are compact, regularly--shaped objects, 
whereas we expect starburst galaxy to look much more irregular (in particular,
if the starburst is triggered by a merger, or if a large amount of dust 
irregularly 
distributed is present in the galaxy). HR10, imaged by HST, is consistently 
characterized by a clearly disturbed morphology (see Dey et al. \cite{dey99}).

For what concerns the total number density of EROs and their link
with passively evolving ellipticals, the works by Cowie et
al. \cite{cowie1} and Hu \& Ridgway \cite{hu} suggest that at most a 
fraction of the present day ellipticals ($\sim$ 10\%) could have its 
progenitors among EROs, but at present the question 
is far from being settled. Thompson et al. \cite{thomp} and Barger et al. 
\cite{barger}, for example, raise this estimate by a factor 4 $\sim$ 5; 
Ben\'{\i}tez et al. \cite{benitez} claim that the density
of luminous galaxies is comparable with the local value up to $z\sim 2$;
Eisenhardt et al. \cite{ei2000}, finally, argue that the fraction of red
galaxies at $z>1$ might be significantly higher than previously
thought, and consistent with a pure luminosity evolution scenario. 
Finally, Daddi et al. \cite{daddi} recently showed that EROs are
strongly clustered and that such a clustering can explain the origin
of the previous discrepant results on the surface density of $z>1$ 
elliptical candidates as due to strong field-to-field variations. 
It has also been noted
that even a small amount of star formation would drive a high--redshift 
elliptical galaxies towards bluer colors, so that it would be missed by
a sample selection based on photometric properties only (for example,
see Schade et al. \cite{schade}).
The problem of identifying high-redshift evolved galaxies, therefore, 
is not restricted to the ERO population alone; in this perspective, 
color--based selection criteria appear insufficient. 
Again, a different diagnostic tool (Franceschini et al. \cite{france}, Schade 
et al. \cite{schade}) is
provided by a quantitative analysis of morphological characteristics.
A local elliptical galaxy is an evolved system from the point of view
of both its stellar population and its internal dynamics; we may presume 
that, for some objects at high-$z$, a residual small star formation activity
(and in general the overall stellar content of the galaxy) could affect 
the global colors but leave the shape of the brightness distribution 
more or less unchanged, so that such galaxies could be easily identified on 
a morphological basis.
Of course this kind of approach requires imaging at high angular resolution, 
such as can only be obtained by space observatories (namely, by the Hubble
Space Telescope -- HST hereafter). 

As a first effort to investigate the morphology of EROs, we present here 
a quantitative analysis carried out on deep HST archive images both in 
the optical red and in the near infrared. 
Our aim is to identify elliptical galaxies using their morphological 
characteristics, and establish their fractional abundance with respect
to the overall ERO population, assuming that their surface brightness 
distributions at $z > 1$ are similar to the ones observed in the local 
universe.
In particular, at the resolution provided by HST, a first classification can 
be performed visually between compact and irregular objects; among the 
former ones, different distributions can then be distinguished by fitting
different models to the data (for example, exponential and de Vaucouleurs 
profiles, typically associated to disk galaxies and ellipticals
respectively).
To this purpose, we have implemented a code for the analysis of the surface 
brightness distributions of such objects as observed by HST, tested 
its accuracy on a large number of simulated galaxies, and applied it
to a sample of 41 EROs.
%In perspective, this exercise can be considered a preliminary test for a more 
%extended study to be carried out on deep HST images (e.g., the Hubble Deep
%Field South), aimed at the identification -- based on morphological criteria --
%of the whole population of high--redshift ellipticals. 

This paper is organized as follows:
we start describing our sample and the data available for every galaxy,
turning afterwards to discuss in detail the techniques developed for
the final steps of the data reduction, and for the data analysis;
the discussion of the resulting parameters and a morphological classification 
of the sample are carried out in Sect. \ref{sect_res} and \ref{sect_disc}; 
the conclusions follow in Sect. \ref{sect_conc}.

\section{The sample}

We selected a sample of candidate high--redshift ellipticals
from a collection of EROs with published optical--infrared colors.
We restricted our choice to the galaxies for which deep images 
in the red or near--infrared photometric bands were available from the HST 
archive -- in particular observed by WFPC2 or NICMOS; roughly speaking, 
for objects at $z \geq 1$ these two instruments map respectively the UV and
optical rest-frame spectral regions of the emitted radiation. 
Four of the sample galaxies are in the Hubble Deep Field South (HDFS), 
and were studied by Ben\'{\i}tez et al. \cite{benitez}.

\begin{figure}
%\vspace{1.5cm}
\resizebox{\hsize}{!}{\includegraphics*[10mm,55mm][190mm,240mm]{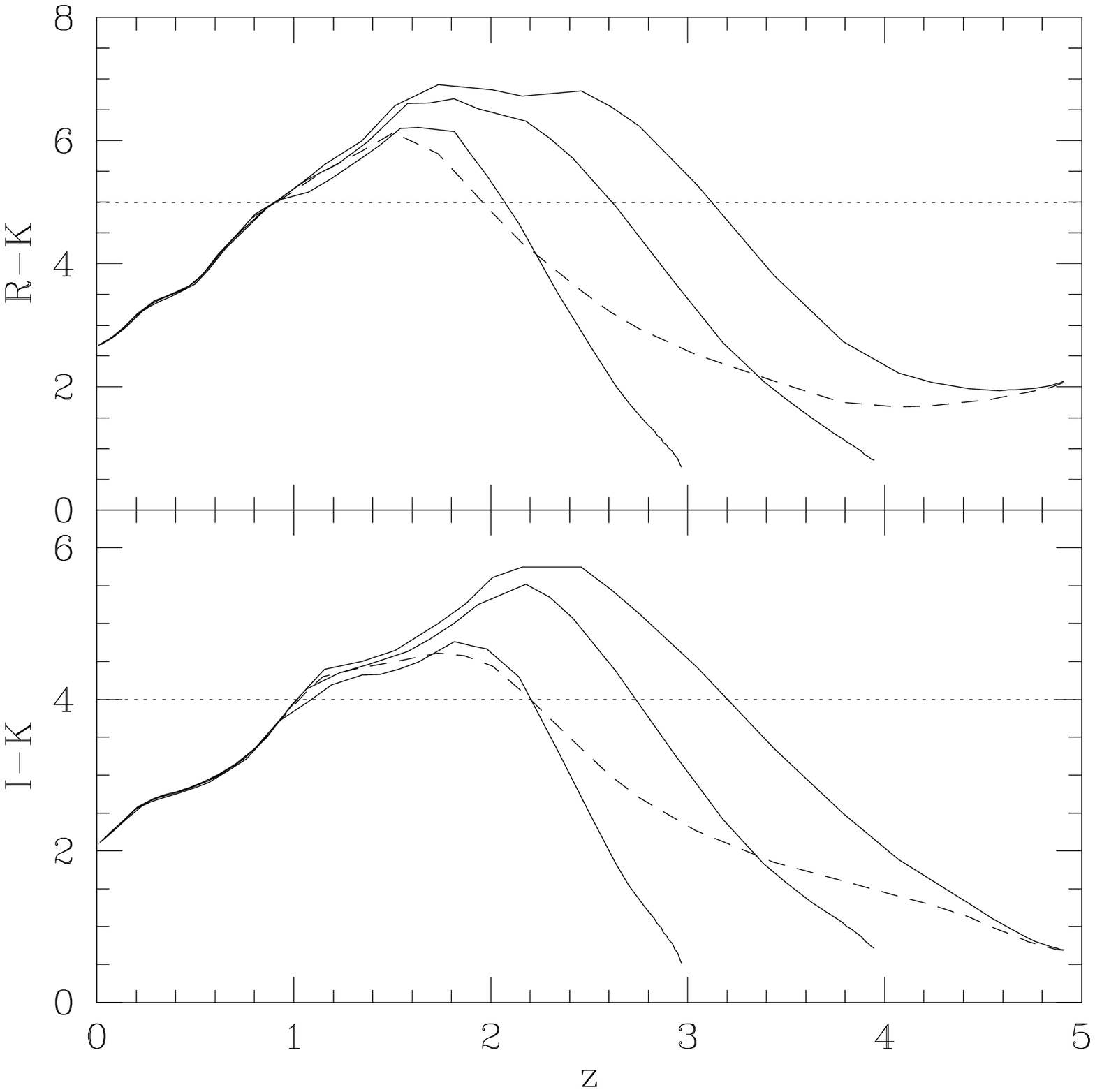}}
\caption{
The curves show the $R-K$ and $I-K$ expected colors for passively evolving
ellipticals. The models assume an exponential decaying star formation
rate (SFR) with e-folding time $\tau$, solar metallicity, Salpeter IMF,
and they are based on the 1997 release of the Bruzual \& Charlot (\cite{bc}) 
spectral synthesis models. The assumed cosmology is 
$\Omega_0=1$ and $H_0$=50 km s$^{-1}$
Mpc$^{-1}$. The solid lines have $\tau=0.1$ Gyr and z$_{\rm formation}=$3,4,5.
The dashed line has $\tau=0.3$ Gyr and z$_{\rm formation}=$5. The dotted
lines at $R-K=5$ and $I-K=4$ show the color thresholds that can be
defined to select elliptical candidates at $z\geq1$. Redder color
thresholds allow the selection of higher-$z$ ellipticals (e.g.
$R-K>6$ or $I-K>4.5$ correspond to $z\geq1.3$).
}
\label{fig_ed}
\end{figure}

All the selected objects have $I-K \geq 4$ and/or $R-K \geq 5$;
such thresholds are appropriate to
select candidate elliptical galaxies at about $z \geq 1$, as shown in Fig. 
\ref{fig_ed}, where we plot the theoretical evolution of the observer--frame 
colors, as a function of $z$, for typical stellar populations that we may 
expect to find in elliptical galaxies. \\
The resulting set comprises 63 galaxies, but 22 of them are 
either hardly visible or not detected at all in the final reduced images 
(see the next section), and have therefore been excluded from the 
following analysis; these galaxies are listed in Table \ref{table:badsample}.
The final sample of 41 EROs is listed instead in Table \ref{table:oksample}, 
together with some relevant information on each object; 
colors and $K$ magnitudes from the literature are reported in 
Table~\ref{table_par}. The $K$--band magnitudes span rather homogeneously
the range between 18 and 21, whereas the typical colors are $5 < R-K < 7$ 
and $4 < I-K < 5.5$.

Clearly, the sample was not selected according to any fixed limit in
total flux 
or surface brightness, but it rather comprises objects observed in different 
passbands and with different sensitivities\footnote{In 
Table \ref{table:oksample} and \ref{table:badsample} we report the limiting 
senstivity for the single frames, measured as the surface brightness (mag 
arcsec$^{-2}$) at a 3--$\sigma$ level on the single pixel; we estimated it 
as $\mu_{\rm lim}= ZP - 2.5\, \log (3\, \sigma_{\rm pix}) + 5 \log s$, where 
$\sigma_{\rm pix}$ is the measured noise on the single pixel and $s$ is the 
image scale in arcsec pixel$^{-1}$.}: as a consequence, it cannot be 
considerd complete at any flux level. On the other hand, the 
selection is based only on the availability of deep HST images, so that
we do not expect any particular bias to be present; we also note that,
in spite of its incompleteness, the size of this sample is 
unprecedented for this class of objects.
Finally, since we are mainly interested in the structural characteristic
of these objects, rather than in their intrinsic photometric properties, 
the lack of information about the redshift of most of the selected 
galaxies (9 spectroscopic and 6 photometric redshifts are available from
the literature) does not represent a major problem. The surface brightness 
distribution of local elliptical galaxies exhibits little shape variation 
from the near ultraviolet to the near infrared so that similar passively
evolving galaxies 
at $z = 1 \sim 1.5$ can be easily identified with our kind of analysis 
and with the photometric bands available. The possible effects of the different 
wavelength coverage of the WFPC2 and NICMOS images are discussed in section 6.3.

\begin{table*}
\begin{flushleft}
\caption{The final sample}
\protect\label{table:oksample}
%\vspace{1.5cm}
\resizebox{\hsize}{!}{\includegraphics*[10mm,60mm][200mm,270mm]{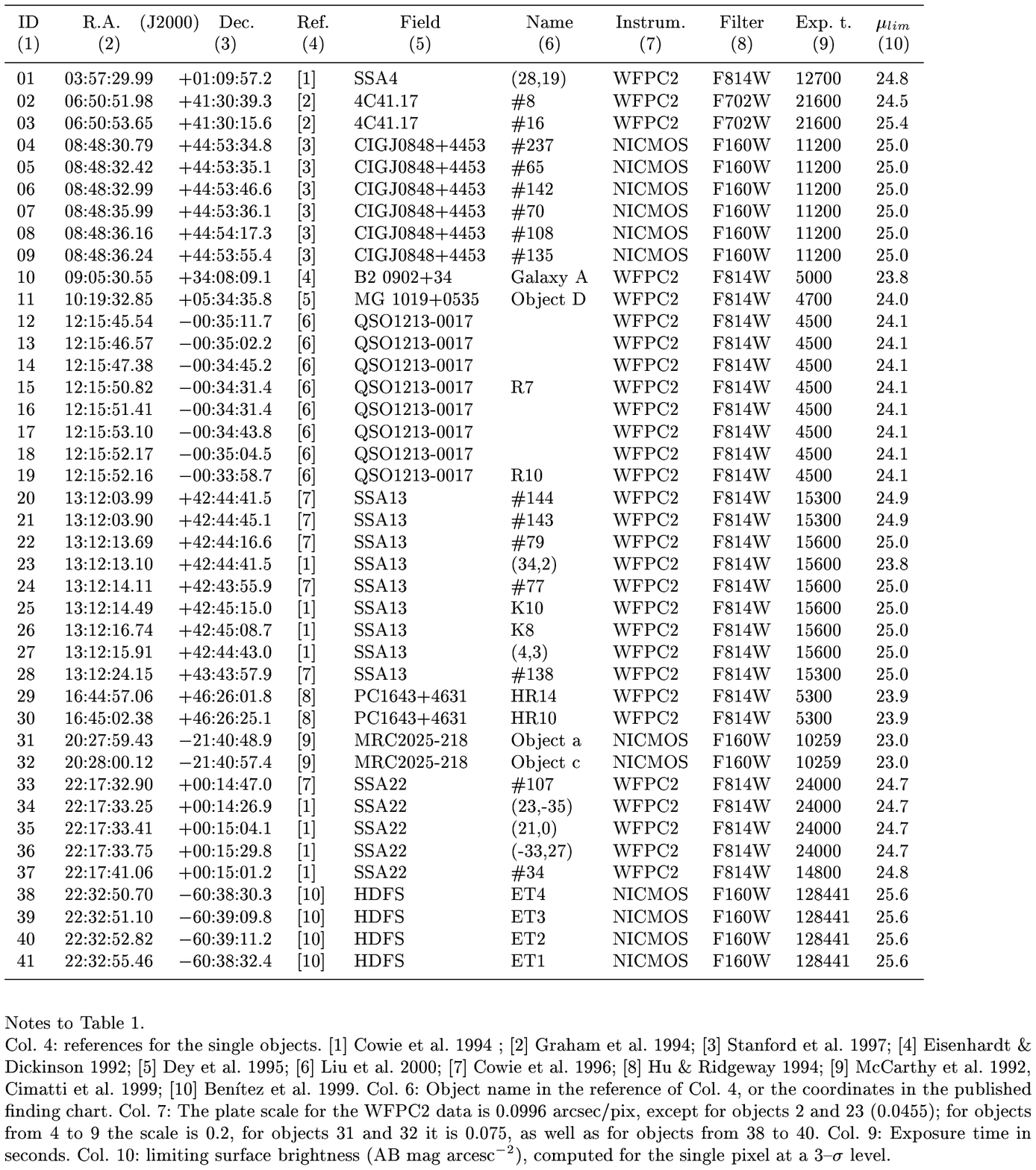}}
\end{flushleft}
\end{table*}
 
\begin{table*}
\begin{flushleft}
\caption{Objects too faint for surface brightness analysis}
\protect\label{table:badsample}
%\vspace{1.5cm}
\resizebox{\hsize}{!}{\includegraphics*[10mm,100mm][200mm,230mm]{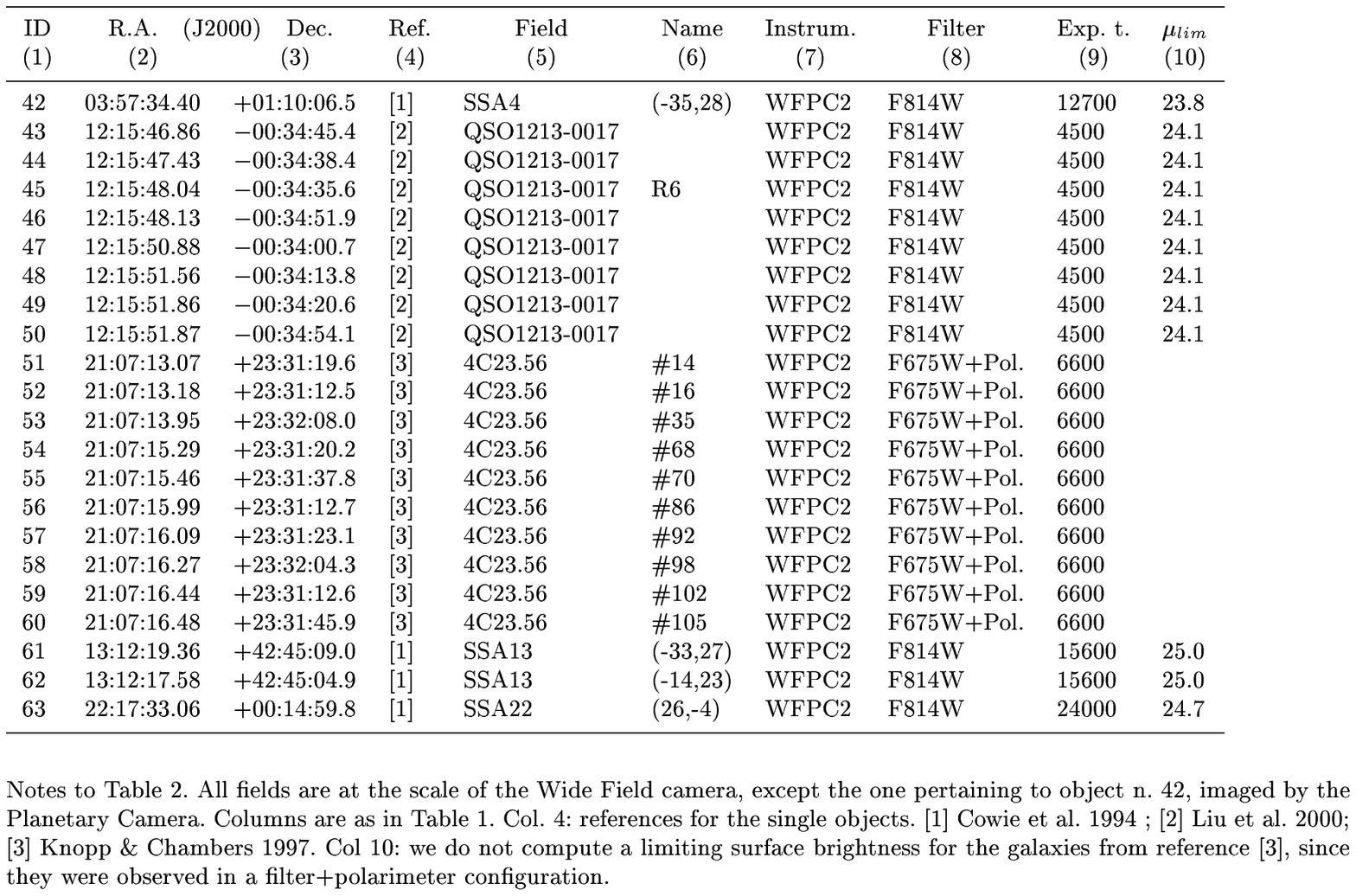}}
\end{flushleft}
\end{table*}
 
\section{Data reduction and analysis \label{sect_dred}}

We have started our analysis from the pipeline--reduced HST datasets 
retrieved directly from the archive. Typically, for 
each field, the observations consist of a sequence of frames slightly 
shifted with respect to each other (shifts range from a few pixels to 
a few tenths), often strongly affected (in particular the WFPC2 ones) by 
cosmic ray hits.
To obtain a single frame per object and to remove at the same time
cosmic rays and residual bad pixels, we have performed a few 
further reduction steps to align and 
combine the images available for each field; a few tests carried out 
on field stars have shown that the procedure adopted to this purpose, 
outlined below, does not significantly degrade the quality of the PSF
in the 
output images, even in the case of relatively large shifts.
All the steps have been performed using routines from the 
IRAF--STSDAS data--reduction packages.

The relative shifts between two different 
exposures of the same field have been evaluated measuring the position of
the peak in the cross--correlation of the two images. 
We have subsequently aligned all the frames using the
DRIZZLE task (Fruchter \& Hook \cite{fruch}), preserving the original 
scale in the output images. We have chosen not to resample the data on 
a smaller scale, since this yields an actual improvement of the spatial 
resolution in the final combined image only if the frames to be combined 
are shifted by 
non integer amounts, and if such shifts sample homogeneously the sub--pixel 
scale, which usually does not happen for our data.
Also, as we will explain in the next section, we compare our data with
model distributions convolved with a theoretical PSF, and the effects 
of the resampling on the PSF are rather difficult to quantify. 
In the end, a better accuracy in the evaluation of the PSF shape more than 
compensates a possible little loss in spatial resolution.
The shifted frames are combined in a following stage,
rather than by DRIZZLE itself, to achieve a more efficient cosmic--ray
rejection. In the case of the HDFS, the public F160W image has been used, 
without any further processing.

Using the ELLIPSE task in IRAF, a radial surface brightness profile has 
been extracted for all the galaxies except one very irregular object 
(number 21 in Table \ref{table:oksample}).
The final images of the sample galaxies are shown in Fig.
\ref{fig_samp} on a logarithmic scale, together with the respective
radial brightness profiles. The map of object 21 is shown in 
Fig.~\ref{fig_dif}: we identify the ERO with the irregular
object in the center of the map, but we do not exclude that the
two close components observed may be part of a single interacting system.

\begin{figure*}
%\vspace{1.5cm}
\resizebox{\hsize}{!}{\includegraphics*[15mm,35mm][200mm,270mm]{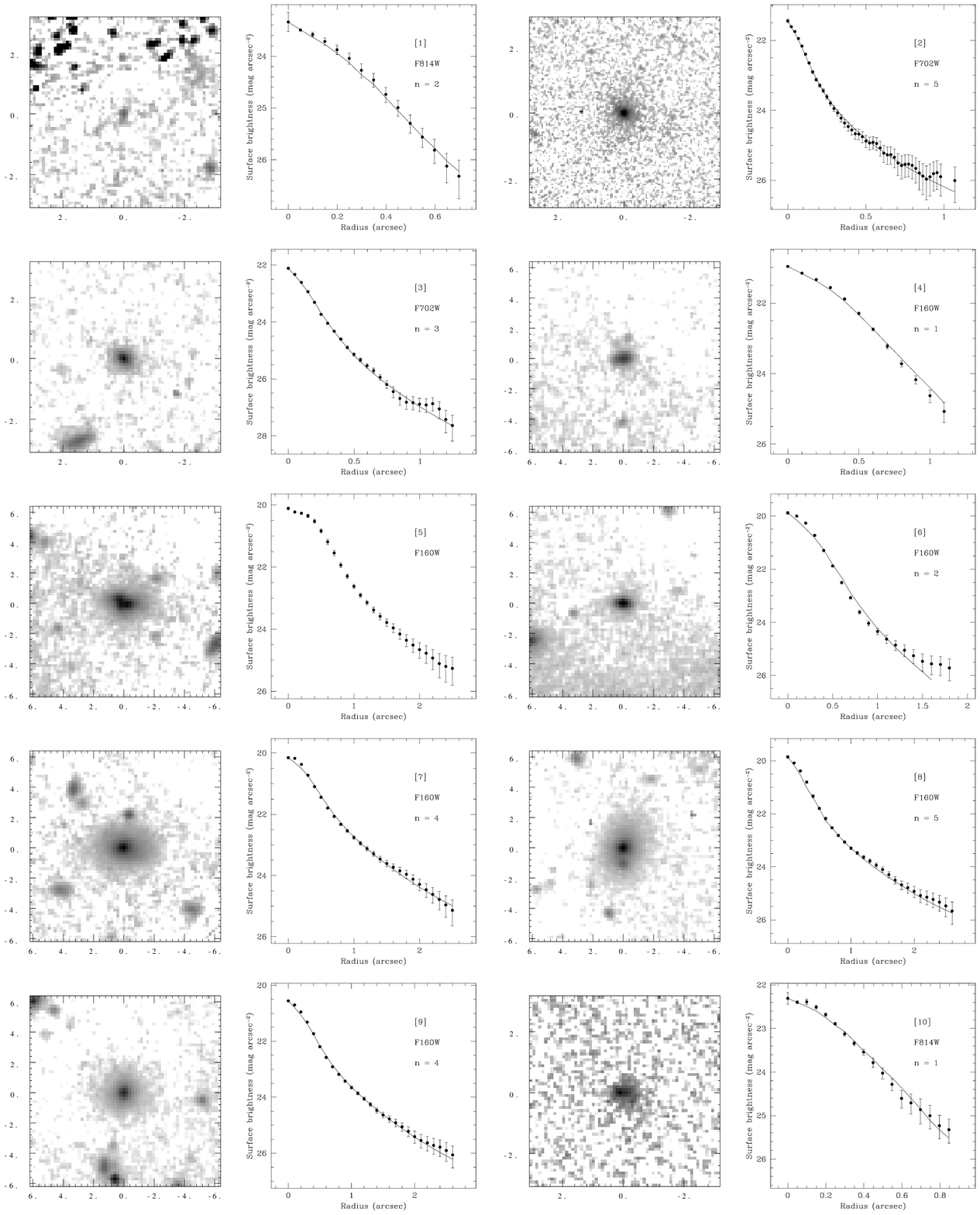}}
\caption{
For each galaxy we show the final image in a logarithmic scale, and a plot
with the radial surface brightness profile.
The orientation of each map is with north up and east to the left, and
the scale on the two axes is in arcsec; the filter is specified 
in the plot panel. Here, the dots and the solid line represent respectively the surface 
brightness profile of the galaxy and of the best fit model;
both profiles are computed as the average intensity of the distributions along 
the same set of elliptical contours.
For objects n. 5 and 30 a satisfactory fit could not be obtained.
Magnitudes are computed in the AB system.
}
\label{fig_samp}
\end{figure*}

\setcounter{figure}{1}
\begin{figure*}
%\vspace{1.5cm}
\resizebox{\hsize}{!}{\includegraphics*[15mm,35mm][200mm,270mm]{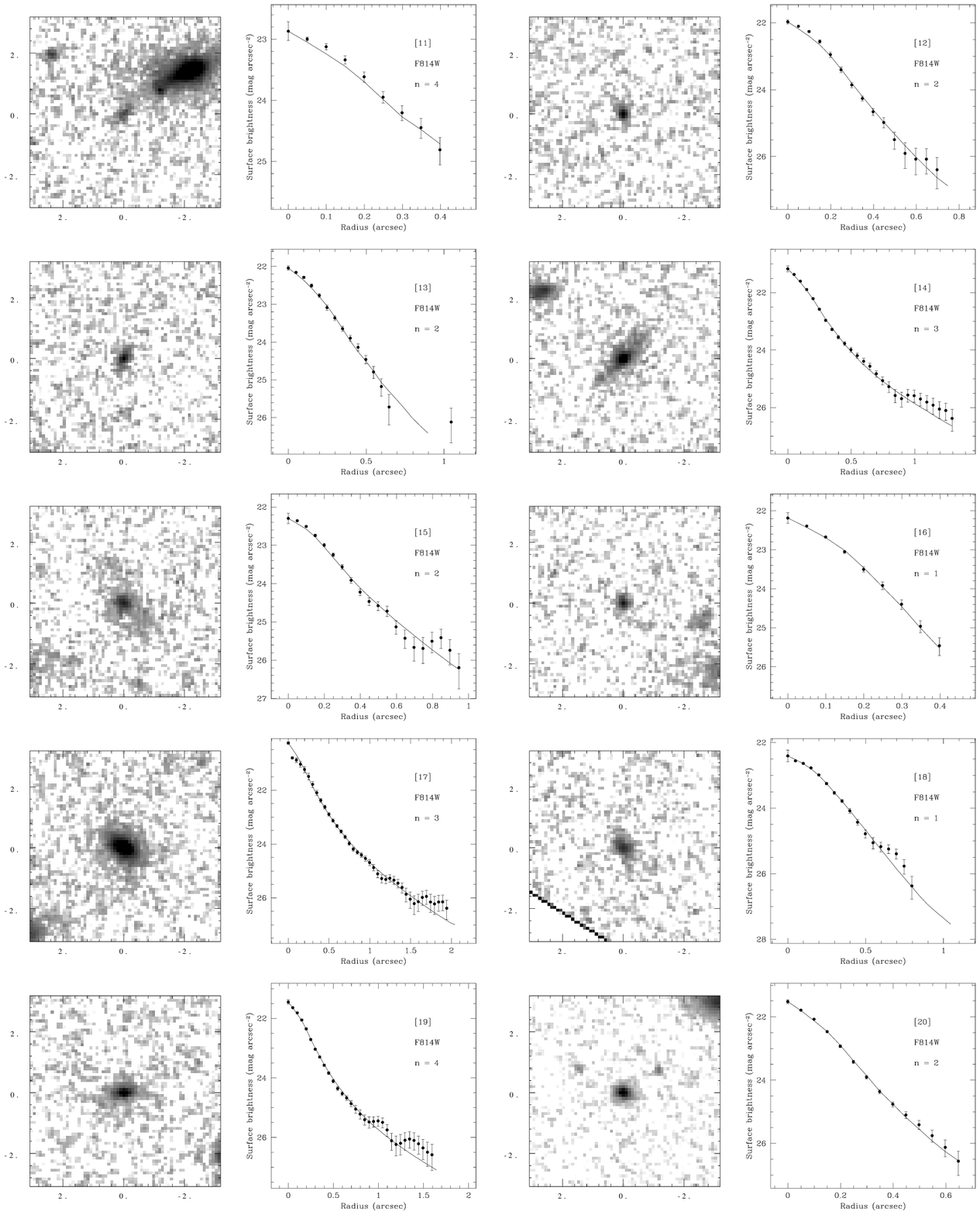}}
\caption{
Continued.
}
\end{figure*}

\setcounter{figure}{1}
\begin{figure*}
%\vspace{1.5cm}
\resizebox{\hsize}{!}{\includegraphics*[15mm,35mm][200mm,270mm]{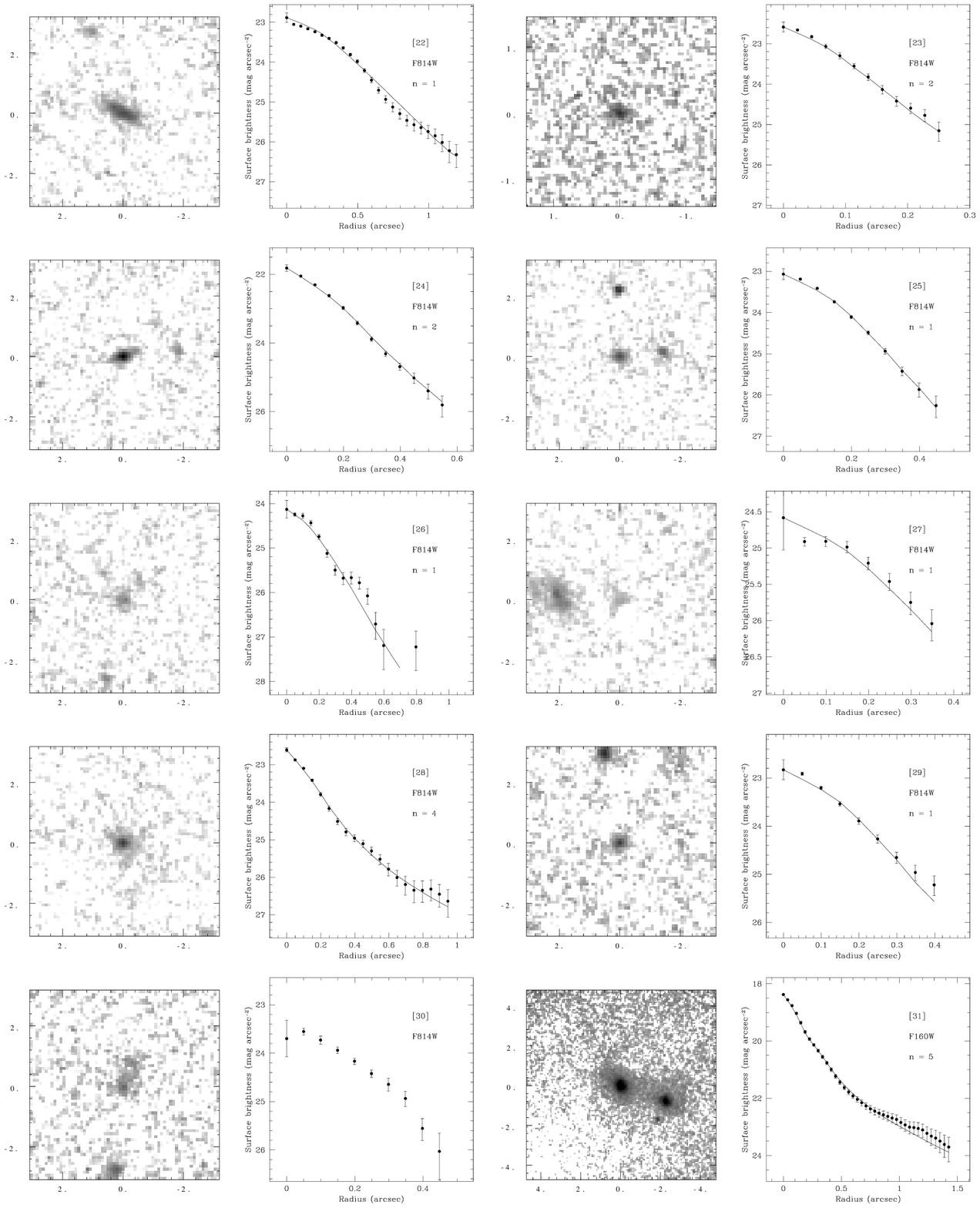}}
\caption{
Continued.
}
\end{figure*}

\setcounter{figure}{1}
\begin{figure*}
%\vspace{1.5cm}
\resizebox{\hsize}{!}{\includegraphics*[15mm,35mm][200mm,270mm]{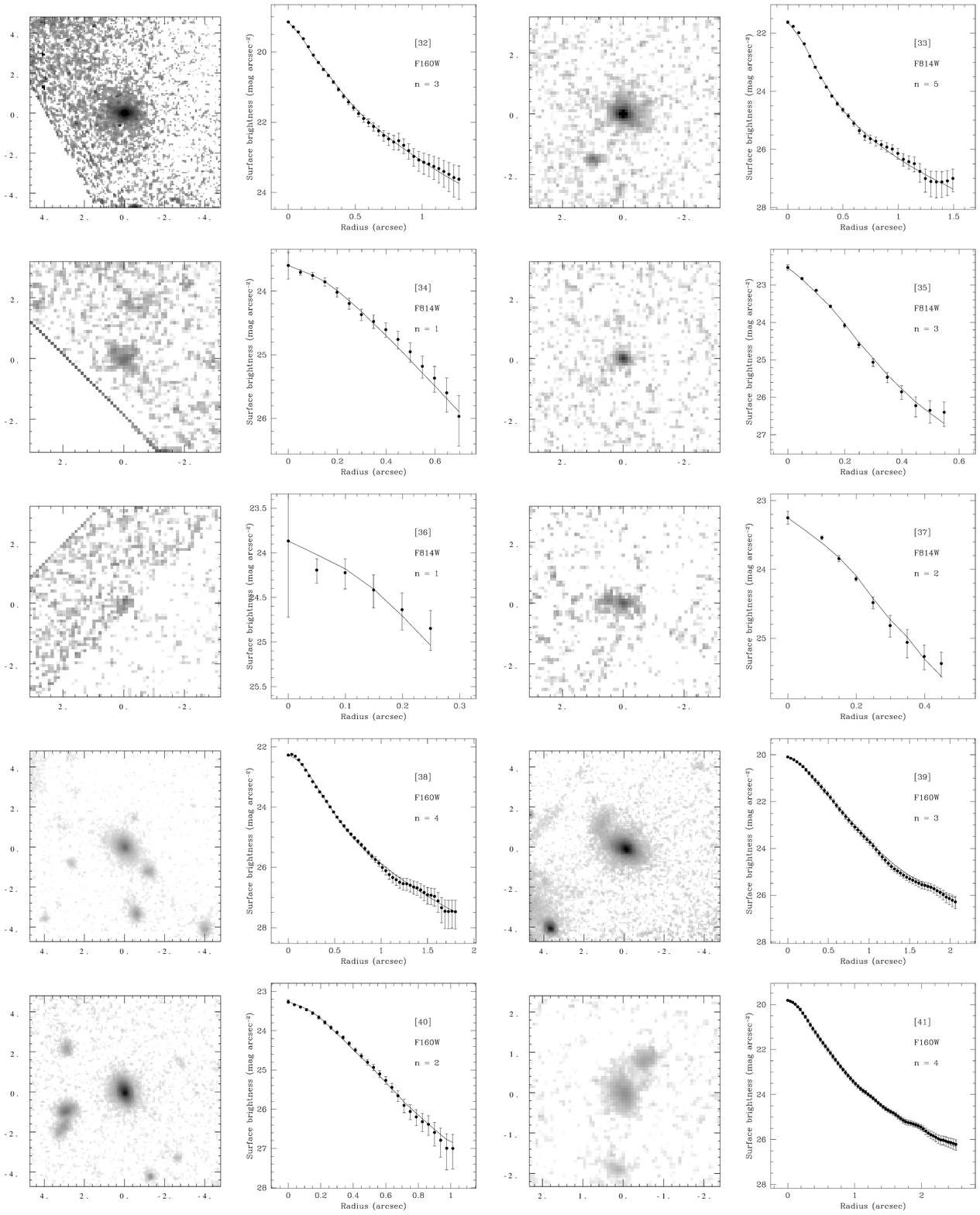}}
\caption{
Continued.
}
\end{figure*}

\begin{figure}
%\vspace{1.5cm}
\resizebox{\hsize}{!}{\includegraphics*[45mm,65mm][165mm,175mm]{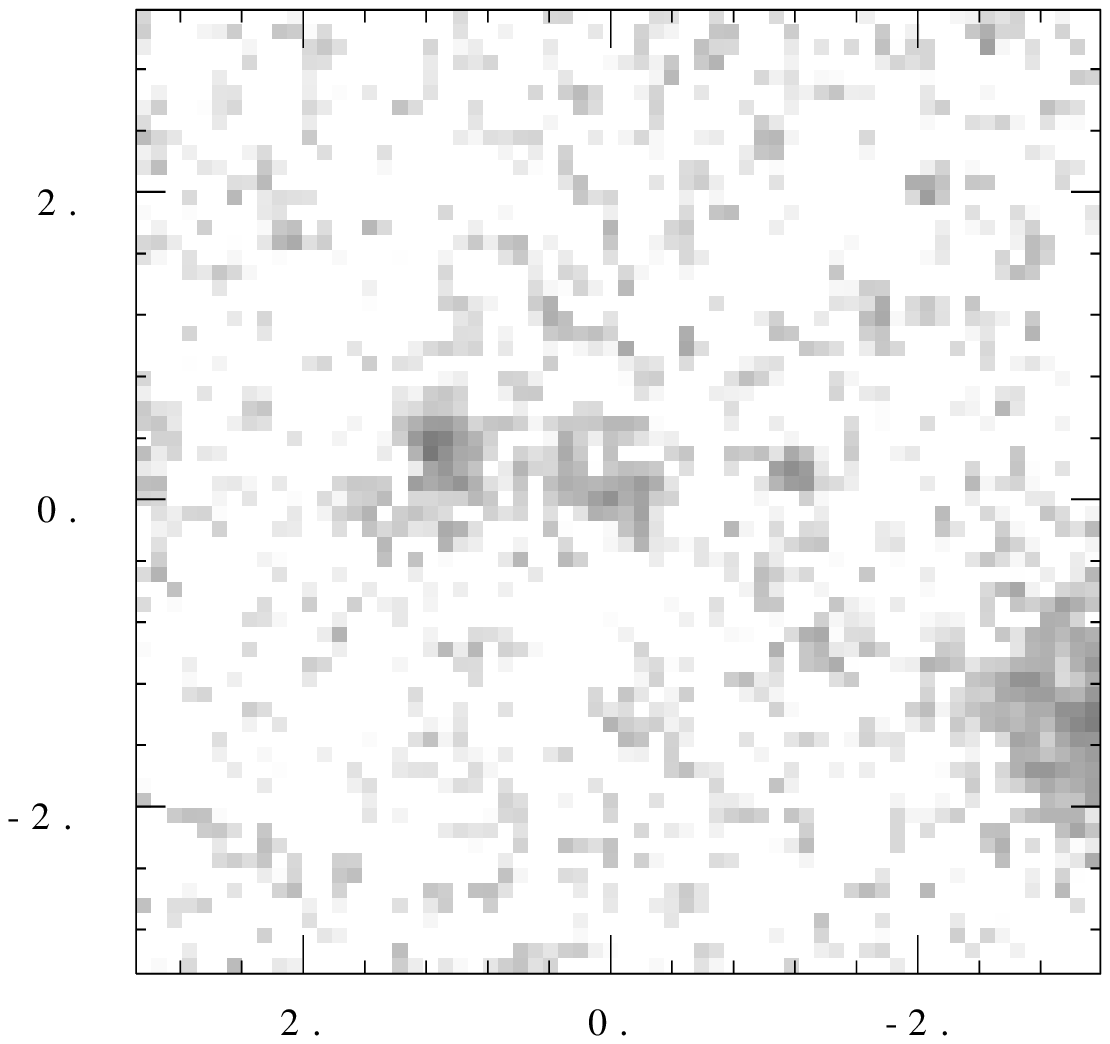}}
\caption{
We identify object n. 21 with the one in the center of the map (the filter 
is F814W). We did not obtain a fit for this galaxy due to its disturbed 
morphology, and to the closeness of a bright companion. 
}
\label{fig_dif}
\end{figure}

\section{Fitting the surface brightness distributions}

A model brightness distribution has been fitted to each object,
using a modified version of the code described in Moriondo et al. 
\cite{moriondo}. Such a code was originally implemented to analyze 
the brightness distribution of nearby spiral galaxies by fitting 
to the data a bi-dimensional two--component model (disk+bulge), convolved 
with a gaussian PSF. 

\subsection{The Point Spread Function}

The main changes introduced to make the fitting code suitable for the 
WFPC2 and NICMOS data concern the convolution of the model galaxy with 
the PSF, which is not axisimmetric and subject to significant changes 
from one point to another in the field. In other words, the gaussian 
approximation is no longer accurate enough, and the PSF needs to be 
represented by a full bi--dimensional image.
The model PSF were computed using the TinyTim code (Krist \& Hook
1997, {\tt http://www.stsci.edu/ftp/software/tinytim/}). Such a
theoretical PSFs proved to be accurate within about 15\%, when compared
with observed stars and considering the average residuals in the inner 
5 $\times$ 5 pixels. The accuracy achieved by TinyTim is not worse than 
what could be obtained using stars in the field; this is mainly due to the
variation of the PSF shape across the field of view 
(differences up to 20\%, even in the central pixels, are easily observed),
coupled to the fact that it is usually difficult for WFPC2 images -- almost 
impossible for the NICMOS ones -- to find a star in the neighbourhood of each 
object. 
A further advantage offered by TinyTim is the possibilty of computing an
oversampled PSF, which greatly improves the accuracy of the model convolution.
This is particularly true in the case of WFPC2 data, where the image scale
undersamples the PSF: a few tests have shown that in this case a good 
convolution of the model can be obtained only if it is performed on a finer 
grid. 
The convolved model is then rebinned to the proper scale and compared to the 
data.
In the case of the HDFS, since the reduced frames have been resampled 
by DRIZZLE, we have chosen to use a PSF derived directly from the stars 
in the field. 

The convolution is performed as the inverse Fourier transform of the
product of the direct transforms of the model and the PSF, using a Fast 
Fourier Transform algorithm.

\subsection{The model}

Because we want to identify elliptical galaxies, and
given the small size of the sample objects, we have considered only 
one-component models with constant apparent ellipticity (i.e. no 
bulge+disk galaxies). The radial trend adopted for the model brightness 
distribution in every fit is a generalized exponential (S\` ersic 
\cite{sersic}): 
$\mu \sim \exp (\alpha_n R^{1/n})$, including the case of an 
exponential distribution ($n=1$) and of a de Vaucouleurs one ($n=4$).
The parameters of each fit are the effective radius $R_e$ and the effective
surface brightness $\mu_e$
of the distribution, as well as its center coordinates; the apparent
ellipticity and position angle are held fixed since they can be 
determined more reliably from the ellipse fitting routine. 
The ``shape index'' $n$ is also fixed, in every fit, to an integer value
ranging from 1 to 6, due to the fact that, at the low signal--to--noise 
ratio ($S/N$) 
typical of our data, the fitting routine is not able to obtain a reliable 
estimate for both $n$ and the other parameters at the same time.
The best value of the shape index $n$ for each galaxy is determined instead
a posteriori, by choosing the least $\chi^2$ resulting from the different
fits. The accuracy of the final best--fit parameters, including $n$, 
has been assesed using a large set of simulated galaxies, and will be 
discussed in the next section.

The fits are performed inside a circular region centered on the galaxy;
its radius is chosen, using the radial brightness profile, as the one
at $S/N = 1$ for the ellipse--averaged intensity.
The background level is estimated on blank sky regions close to the
source; its uncertainty turns out to be dominated, in most cases, by
fluctuations on scales of order 10 pixels or more, due to a non--perfect 
image flattening. 
% Say about error on bg?

\section{Simulations of faint galaxies
\protect\label{sec_sim}}

To establish the accuracy that can be attained by the fits for all the 
relevant parameters, we have
tested our code on a large set of simulated galaxies. We have chosen model
distributions spanning the same range in effective radius and total 
flux -- in terms of instrumental units -- as our sample galaxies,
and a range of values (from 1 to 5) for the shape index.
Since most of the objects considered have been imaged with the wide field 
camera of WFPC2, we have convolved the test distributions with a typical 
wide--field PSF. In this case, the plate--scale is the one which mostly 
undersamples the PSF itself, yielding the worst conditions for the 
retrieval of the correct parameters' value. We expect therefore 
the uncertainties derived from our simulations to be basically correct 
for the WFPC2, and conservative estimates for 
the other types of data: a few simulations carried out with the 
characteristic sampling of the Planetary Camera and of NICMOS show that 
this is indeed the case.

Noise has been added to the simulations at a level typical for our data, 
both on the 
pixel scale and at smaller spatial frequencies. The resulting images have
been finally fitted using 
our code, allowing for some error in the estimate of the PSF and using
different trial values for $n$, as in the case of the real galaxies. 

\begin{figure*}
%\vspace{1.5cm}
\resizebox{\hsize}{!}{\includegraphics*[5mm,50mm][190mm,240mm]{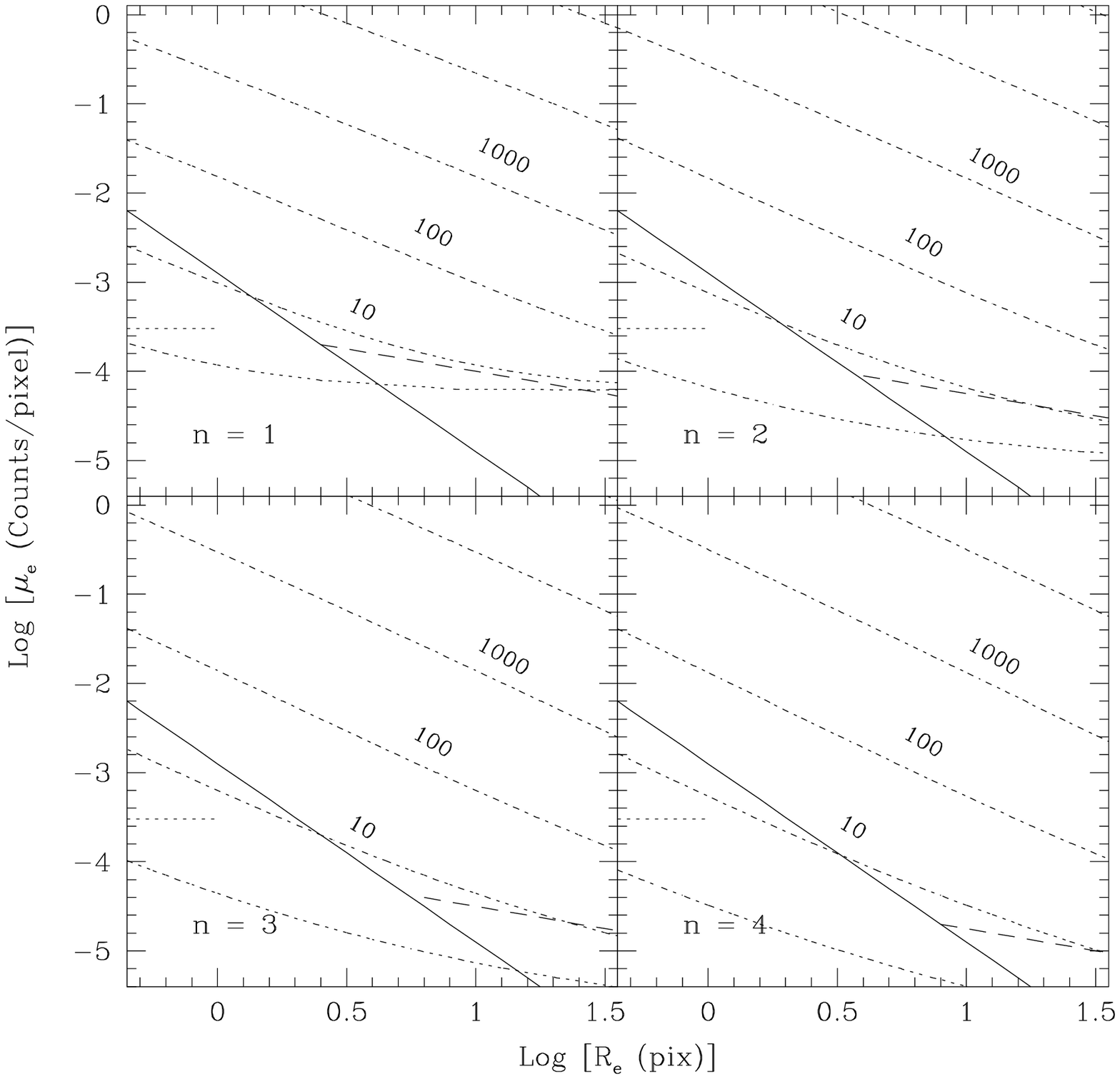}}
\caption{
The region covered by the simulations, and the estimated $S/N$
ratio. The different panels show the results for different values of $n$, 
from 1 to 4.
Each panel represents the $R_e$--$\mu_e$ plane in instrumental units
(pixels and counts per pixel). 
The region covered by our simulations, including the area occupied
by the sample EROs, lies above the solid line; its slope 
is at constant total flux.
The level marked by the dotted horizontal segment is the noise level of the
background at 1 $\sigma$, whereas the dashed lines in the lower right
corner represent the detection limits for the $n=1$, 2, 3, 4 distributions
in the respective panels.
The dotted contours map the estimated $S/N$ at the values indicated 
by the labels.  
}
\label{fig_snr}
\end{figure*}

In Fig.~\ref{fig_snr} we show the region covered by the synthetic objects 
in the $R_e$--$\mu_e$ plane, represented in instrumental units (pixels and 
counts per pixel respectively), one panel for each value of $n$, from 1 to 4. 
The dotted contours map the $S/N$, evaluated theoretically, considering the 
total flux inside the isophotal radius at $S/N = 1$ on the single pixel.
These values do not account for the effect of the
PSF, and are therefore less reliable at low surface 
brightness levels and for sizes of the order of the PSF width
(say, $\log (R_e) \sim 0$). For this reason, the detection 
limits for our data have been deduced a posteriori from the simulations, 
and are represented by the dashed lines in the lower right corners of the 
plots. 
% below the lines the galaxies are completely hidden by the background
% noise. 
Such limits appear to depend on $n$; this is partly explained by the
fact that, at fixed effective radius and
surface brightness, there is a slight increase of $S/N$ with increasing $n$.
However, the
greatest contribution (about 80\%) is due to the steeper central peak that
characterizes the large--$n$ distributions, making them more visible in
the background noise.

\subsection{Disentangling the $n=1$ and $n=4$ distributions}

A first check can be carried out
assuming that all elliptical galaxies are characterized by a de Vaucouleurs
brightness distribution ($n$ = 4); we can consider this as a first order
classification, since nearby ellipticals show in fact a variety of
shapes, that can be quantified by different values of the exponential index
$n$ ranging from about 2 to 10 and higher (see for example Caon
et al. \cite{caon}, or Khosroshahi et al. \cite{khos}).
If we restrict our test to the simulations with $n=1$ and $n=4$ only,
we find that the correct value of $n$ can be retrieved, on the basis of a
$\chi^2$ estimator, for the whole parameter space explored. Such a
simplified classification, therefore, is possible for all of our objects.

\subsection{Systematic errors}

\begin{figure}
%\vspace{1.5cm}
\resizebox{\hsize}{!}{\includegraphics*[5mm,50mm][195mm,240mm]{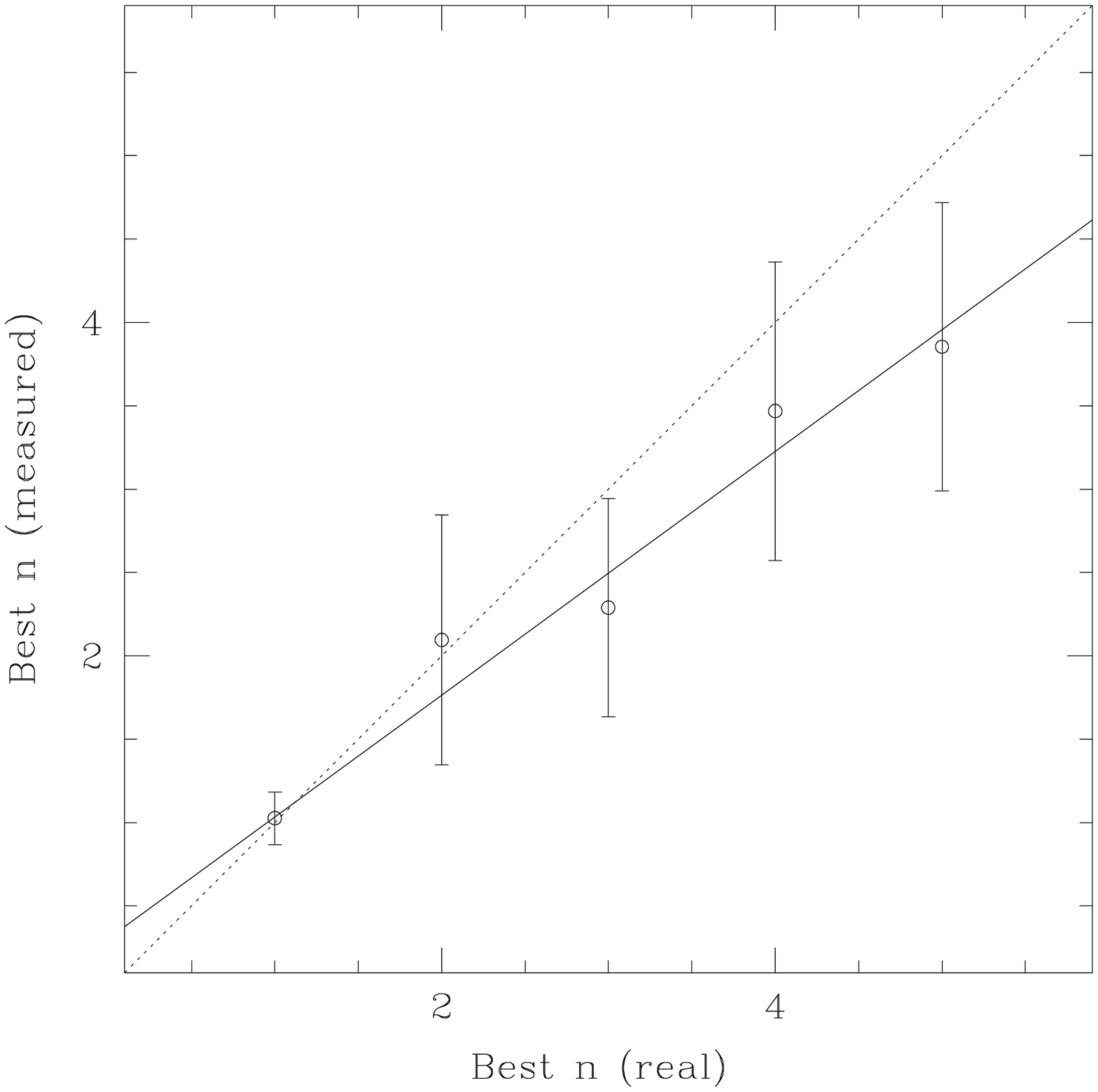}}
\caption{
The trend of measured vs. real $n$, for the simulated galaxies. 
$n$ values larger than 2 tend to be underestimated.
}
\label{fig_cn}
\end{figure}

In the following step we assigned to every galaxy its best--fit $n$ value,
choosing from 1 to 6, and compare the derived parameters with the original 
ones. We can start our analysis of the results by looking for
systematic trends. Starting with our estimates for the index $n$,
if we plot the average measured values vs. the true ones (Fig. \ref{fig_cn}),
we actually observe a tendence for the higher $n$'s to be underestimated;
in particular we find 
\begin{equation} 
\protect\label{eq_cn}
n_{true} = (1.37 \pm 0.30) n_{true} + (-0.41 \pm 0.35) \; \; .
\end {equation}
This effect depends only slightly on the choice of the PSF; it is probably
due to the fact that, increasing $n$ at fixed total flux, the peak 
of the distribution gets sharper, and its wings fainter and wider, but 
these differences tend to be concealed by the effect of the PSF on one 
hand, and by the backgroud noise on the other.

Similar trends can be investigated also for $R_e$ and 
$\mu_e$: we find that, for both quantities, small
values (i.e. those approaching respectively the pixel scale and the 
background noise level) tend to be slightly overestimated, and large values 
tend to
be underestimated. The behaviour is very similar for all the $n$ values,
so that we can adopt average corrections:
\begin{equation}
\protect\label{eq_re}
\log (R_{true}) = (1.15 \pm 0.05) \log(R_{obs}) + (-0.09 \pm 0.03) \; \; .
\end {equation}
\begin{equation}
\protect\label{eq_br}
\mu_{true} = (1.11 \pm 0.03) \mu_{obs} + (0.19 \pm 0.08) \; \; .
\end {equation}
Since these latter corrections are significant at a 3 $\sigma$ level,
whereas Eq. \ref{eq_cn} is significant only at 1 $\sigma$,
and since applying Eq. \ref{eq_cn} to the estimated $n$'s would lead 
to non-integer values for this parameter, we choose to correct only 
$R_e$ and $\mu_e$ and leave $n$ unchanged, keeping in mind 
that $n$ values greater than two might be somewhat underestimated.

\subsection{Mapping the parameter space}

\begin{figure*}
%\vspace{1.5cm}
\resizebox{\hsize}{!}{\includegraphics*[5mm,50mm][190mm,240mm]{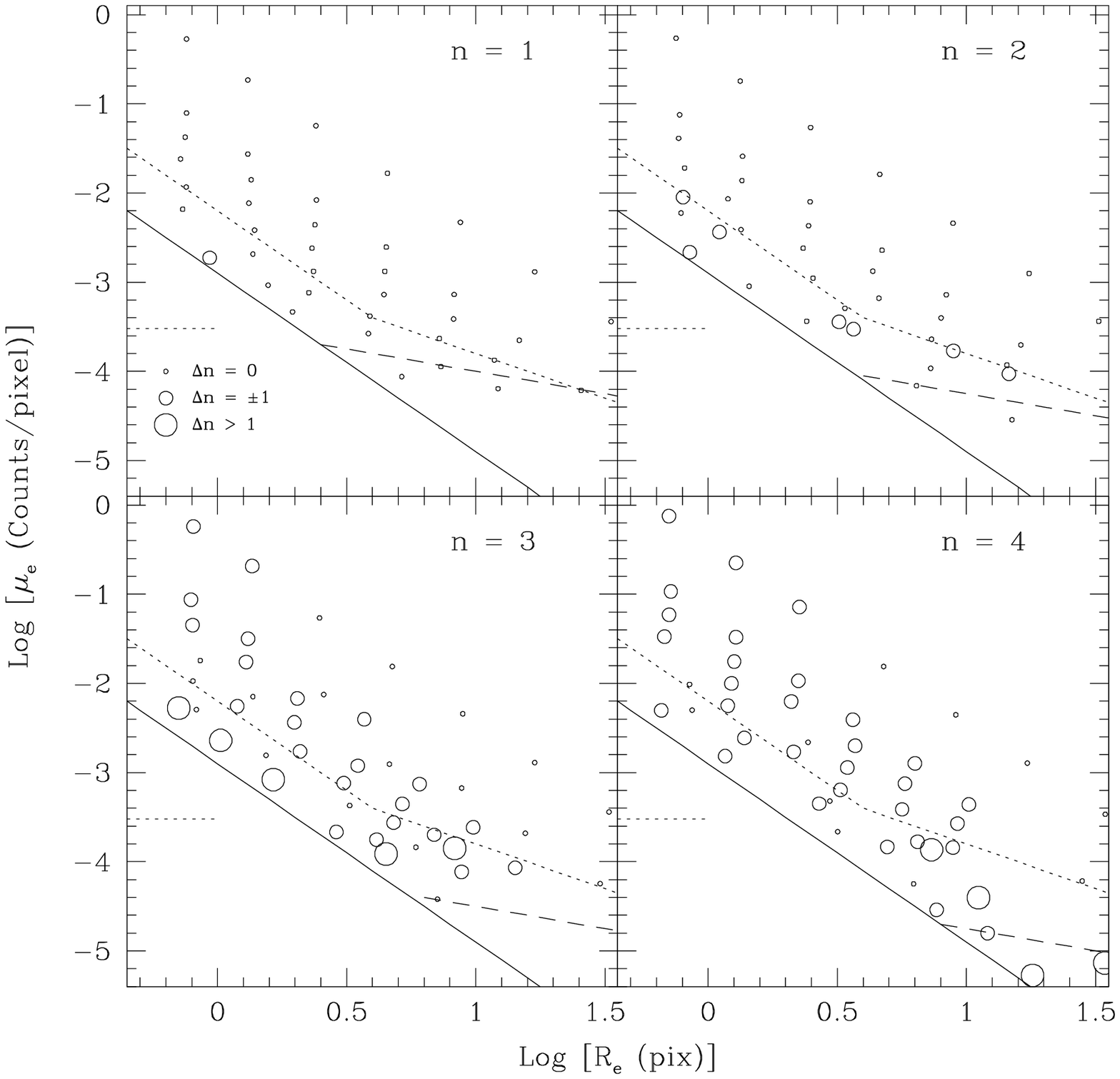}}
\caption{
Accuracy in the estimates of $n$. The plot is similar to the one in Fig.
\ref{fig_snr}. 
In each panel the dots are placed at the estimated location of the simulated
galaxies, with their size quantifying the accuracy
in the estimate of $n$: the small dots correspond to the correct value, the
medium--size ones imply an error of $\pm 1$, the large ones an error
grater then 1. In the region above the dotted line, we can relibly distinguish
between $n=1$ and $n > 1$ distributions.
}
\label{fig_chi}
\end{figure*}

We turn now to examine how accurately the relevant parameters are
retrieved in the various regions of the parameters' space, 
starting with the shape index $n$.
Fig.~\ref{fig_chi} shows again the $R_e$--$\mu_e$ plane; in this plot
the dots in each panel represent the estimated location of our simulated 
galaxies for the different values of $n$.
The accuracy with which $n$ can be retrieved -- without applying 
Eq. \ref{eq_cn} -- is quantified by the size
of each dot, as explained in the caption.

We find that the correct value of $n$ is retrieved in most cases for the
$n=1$ and $n=2$ models; the error for the $n=3$ and $n=4$
ones is more typically 1 in large portions of the plane, partly due to the 
systematic effect described previously. As a consequence, exponentials are
almost always recognized as such, so that if the best fit is for 
$n \neq 1$, the distribution is certainly non--exponential.
As expected, for all values of $n$, low flux and low surface
brightness objects tend to be affected by larger errors.
The main conclusion, however, is that relying on these
results we can define a region (the one above the dotted line)
where exponential distributions can be reliably distinguished from the
others: this is the locus where both exponentials and $n=2$
distributions are recognized as such, and larger $n$ distributions are
affected at most by an error of 1.
A comparison with Fig. \ref{fig_snr} shows that the limit roughly
spans $S/N$ values between 10 and 80.
We have checked this result using the theoretical approach described in Avni
\cite{avni}: when one or more parameters are evaluated via a $\chi^2$ minimization,
the method allows to assign a confidence level to each parameter relying 
on the variations of the $\chi^2$ around the minimum in the parameter space.
Although the computations are exact only in the case of linear fits, 
the method provides anyway a useful check on our findings; indeed, we find that
our estimates for the uncertainty of $n$ are broadly consistent with the
ones evaluated theoretically for a 90\% confidence level.
In particular, the Avni method confirms that in the area above the dotted line, 
exponential and non--exponential distributions can be reliably distinguished.

A mapping of the parameter space, analogous to the one plotted in 
Fig. \ref{fig_chi}, has been produced also to estimate the uncertainties on
$R_e$ and $\mu_e$, corrected according to Eqs. \ref{eq_re} and 
\ref{eq_br}.
An intersting result is that the derived errors are relatively independent 
of the estimate of $n$,
in the sense that a wrong estimate of the shape index does not necessarily
mean larger errors for $R_e$ and $\mu_e$. Most likely, whereas the choice
of $n$ is influenced mainly by the accuracy of the PSF, the estimates
of $R_e$ and $\mu_e$ are more strictly related to the quality of the fit
a whole; in other words, if a wrong $n$ may compensate for the effect of a
wrong PSF, a good estimate for $R_e$ and $\mu_e$ can be achieved anyway,
as long as the quality of the fit is good.

For what concerns the values of the ellipticity and position angle (that
are fixed a priori), we find that the typical errors associated to their
estimates do not affect significantly the accuracy of the output
parameters (center coordinates, effective radius and surface brightness),
nor the choice of the best $n$ value.
We estimate the typical errors on the ellipticity to be around 0.1, 
and from 5 to 10 degrees for the position angles.
The center coordinates are usually determined with great accuracy ($\sim
0.3$ pixels): due to the asymmetries in the psf, this is better
than what can be achieved by fitting an ellipse--averaged profile
to the galaxies.

To summarize, we have tested our fitting code on a large set of
simulated galaxies, and assessed the accuracy that can be attained 
for the various relevant parameters in the region of the parameters' space 
covered by the real data. In the next section, the results presented 
so far will be used to estimate the errors on the parameters derived 
for each galaxy.

\section{Results \label{sect_res}}

\subsection{Irregular and compact objects}

The first classification that can be carried out for our sample is
between compact, isolated objects, and galaxies that are clearly undergoing
an interaction or exhibit an irregular, diffuse shape. The easiest way 
to do this is by visual inspection, since all our objects are well resolved.
We find 35 compact galaxies out of 41, or 85\% of the sample.
The ERO HR10 (Hu \& Ridgway \cite{hu}; {Graham \& Dey \cite{grdey})
is included in the subsample of irregular objects.
We have attempted to recognize in each of these irregular objects
a brighter component, such as could be expected in a merging system,
and obtain a fit of its brightness distribution 
after a proper masking of the surrounding areas. 
This was not possible for object 21, object 5, and object 30 (HR 10): 
in the case of object 21, as we mentioned in Sect. \ref{sect_dred}, the 
surface brightness distribution is too diffuse to isolate a major component; 
object 5 is compact, but its nucleus
has a rather irregular shape, probably due to the superposition
of two or more close and equally bright components; in object 30, a
brighter component can be easily recognized, but we did not obtain a 
satisfactory fit to its brightness distribution.
A radial brightness profile was anyway extracted for object 5 and object 
30, as shown in Fig. \ref{fig_samp}.
A few more galaxies are close to other objects that, however, are not
disturbing their morphology: in these cases a proper masking was also
applied, as indicated in Table \ref{table_par}, to avoid any problem with 
the fitting procedure.

\subsection{Structural parameters}

%Although the fit is not performed on the points shown in
%the plot, but rather on the whole bi-dimensional brightness distribution, 
%in most cases there is good agreement between model and data; as expected,
%the worst results are obtained for the objects classified as
%interacting/irregular.
 
\begin{table*}
\begin{flushleft}
\caption{Photometric and structural parameters}
\protect\label{table_par}
%\vspace{1.5cm}
\resizebox{\hsize}{!}{\includegraphics*[10mm,50mm][200mm,280mm]{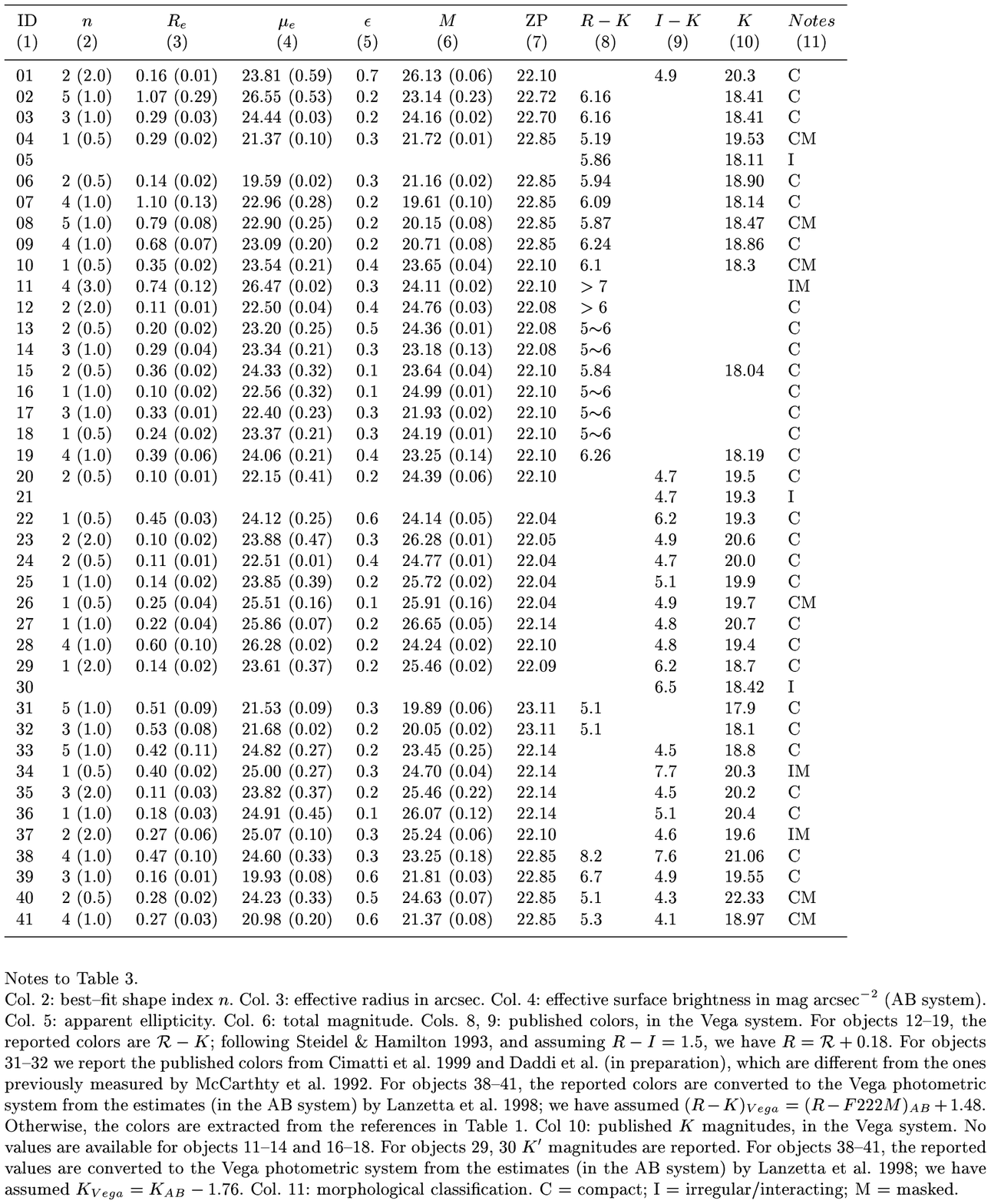}}
\end{flushleft}
\end{table*}

% A further, more detailed classification we are interested in, is between 
% elliptical galaxies
% and non-elliptical ones, which translates -- in the case of compact,
% isolated objects -- into a separation between non--exponential
% and exponential distributions.

The best-fit profiles are plotted in Fig. \ref{fig_samp},
the best--fit parameters for each object are shown in Table
\ref{table_par}, while 
Fig. \ref{fig_galps} shows the estimated location in the $R_e$--$\mu_e$ 
plane of all our sample galaxies with 
different symbols for each instrumental configuration. 
The encircled symbols correspond to the 3 galaxies classified as 
irregular/interacting for which we could obtain a fit, after 
masking the lower flux companion: since for these objects we consider 
only a fraction of the total flux, they are typically 
placed in the lower part of the plot.
Effective radii and surface brightnesses are plotted in instrumental units
(pixels and counts/pixel),
to allow a direct comparison with Fig. \ref{fig_chi}. 
% \ref{fig_re}, and \ref{fig_br}. 
The $y$ coordinates of the data points, however, are not exactly
the values determined by the fit, since each galaxy has been scaled to 
the noise level adopted
for the simulations by applying a proper shift along the brightness axis.
The uncertainties listed in Table~\ref{table_par} are evaluated by
interpolating the results from the simulations at the locations of the 
real galaxies in the parameters' space; we report 1--$\sigma$ errors for
all the parameters except $n$: when its integer value is retrived correctly,
we assign to this quantity a formal error of 0.5, otherwise the integer 
value reported corresponds to the largest possible error.
As we mentioned in the previous section, we have checked -- using a theoretical
approach -- that the estimated errors for $n$ roughly correspond to a 90\% confidence level.
In a few cases (7 out of 38, 2 of which classified as irregular) the thoretical estimate 
exceeds the one derived from the simulations; for these galaxies the uncertainties reported 
are the thoretical ones.

The 4 HDFS galaxies (38--41) were studied by Ben\'{\i}tez et al.
\cite{benitez} with a technique very similar to ours (a best fit 
to the brightness profiles with a de Vaucouleurs' law), so that
their and our results can be easily compared.
We find that a de Vaucouleurs law is the best fit to the data for 
two of these galaxies ($n=4 \pm 1$), the other two being best represented 
by an $n=2 \pm 0.5$ and $n=3 \pm 1$ profile respectively.
For what concerns the integrated 
fluxes, the average difference is 0.07 $\pm$ 0.05 magnitudes,
whereas our effective radii are 0.84 $\pm$ 0.25 times the 
ones by Ben\'{\i}tez et al., on average. We conclude that the differences
between the two works are not relevant, and characterized by only a 
modest scatter in the measured quantities. Excellent agreement is then
found, in the case of object 39, with the results by Stiavelli et al.
\cite{stia}: although our best fit for this galaxy is with $n=3$,
both our effective radius and our total flux are coincident 
with the Stiavelli et al. values, derived adopting a de Vaucouleurs' 
distribution.

Figure~\ref{fig_arcps} shows again the $\mu_e$--$R_e$ plane 
in standard units (arcsec, mag arcsec$^{-2}$), with a different symbol for
each filter; the dotted line represents the slope of constant flux,
at fixed shape index $n$. The plot illustrates the limits in size and 
surface brightness of the sample in the HST filters.

\begin{figure}
%\vspace{1.5cm}
\resizebox{\hsize}{!}{\includegraphics*[5mm,50mm][195mm,240mm]{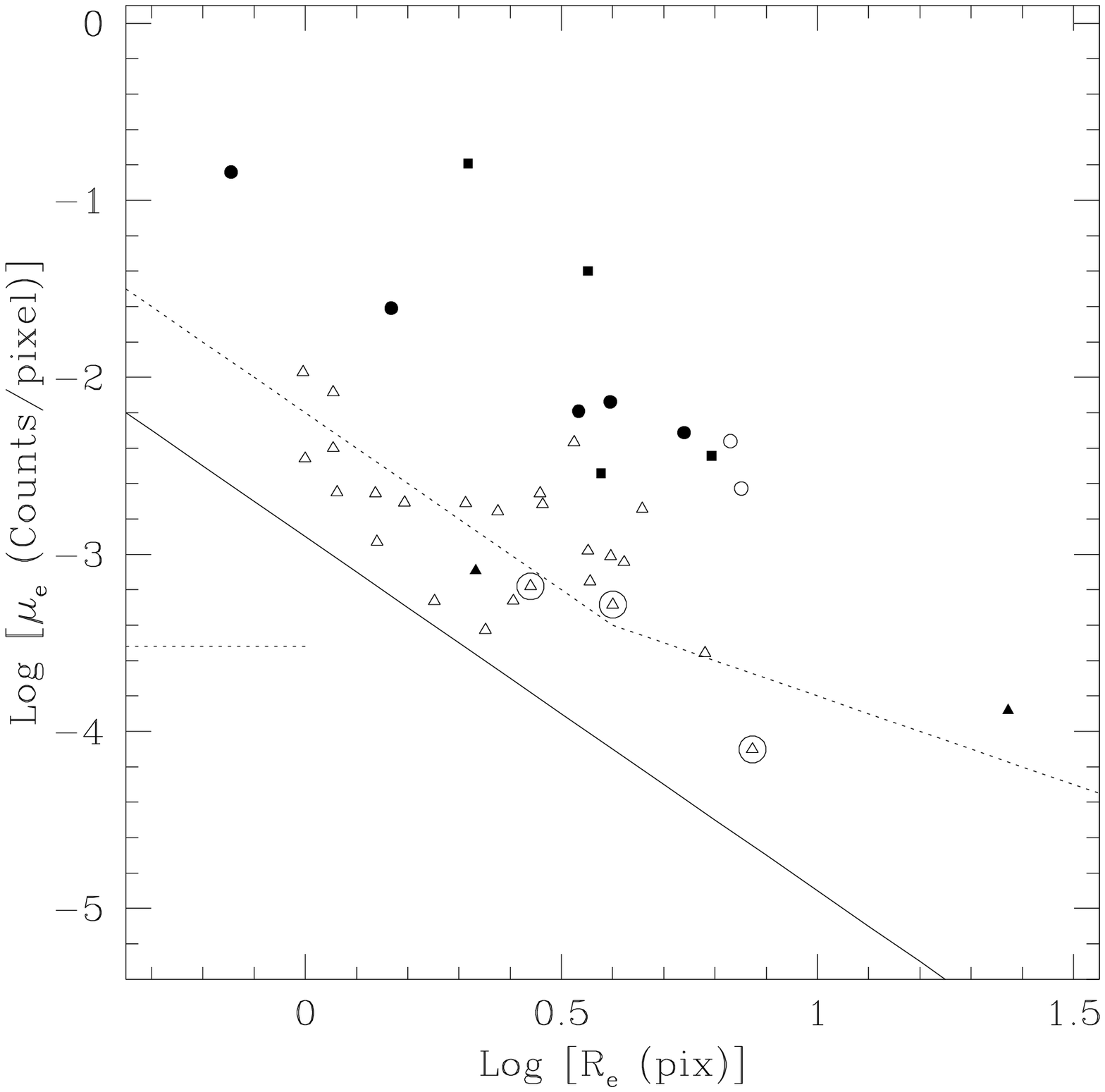}}
\caption{
The location of our sample galaxies in the $\mu_e$--$R_e$ plane;
the units are counts per pixel and pixels respectively.
Different symbols correspond to different
instruments and/or detectors: filled circles for NICMOS camera 3; open circles
for NICMOS camera 2; filled squares for the HDFS galaxies; open triangles
for Wide Field data; filled triangles for Planetary camera data.
The encircled points are those classified as irregular/interacting
in Table~\ref{table_par}.
The other symbols are as in Fig.~\ref{fig_chi}. 
% \ref{fig_re}, and \ref{fig_br}. 
}
\label{fig_galps}
\end{figure}
 
\begin{figure}
%\vspace{1.5cm}
\resizebox{\hsize}{!}{\includegraphics*[5mm,50mm][195mm,240mm]{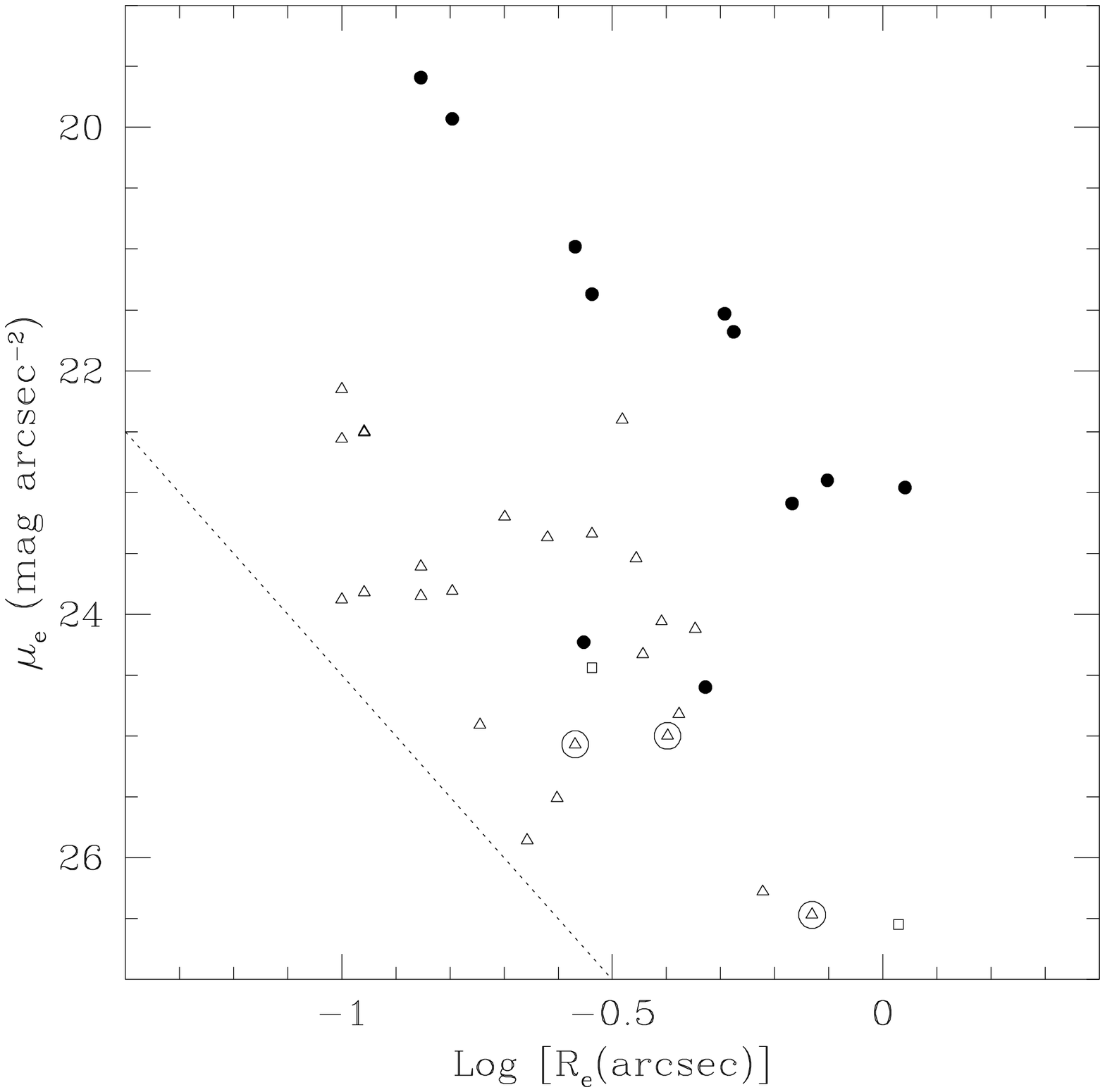}}
\caption{
The location of our sample galaxies in the $\mu_e$--$R_e$ plane;
the units are mag arcsec$^{-2}$ and arcsec respectively.
In this plot different symbols for the data points correspond to different 
filters: filled circles for F160W, open triangles for F814W, open
squares for F702W.
The encircled points are those classified as irregular/interacting
in Table~\ref{table_par}.
The slope of the dotted line is at constant flux.
}
\label{fig_arcps}
\end{figure}
 
\subsection{The shape index $n$ and the fraction of ellipticals}

We evaluated previously, through visual inspection, the fraction of 
irregular objects, concluding that they constitute only a minority of 
our ERO sample. For what concerns the shape of the best--fit distributions, 
we performed two types of classifications. 

A first order classification was performed by comparing the results 
assuming that each galaxy can be properly described by either an
exponential distribution ($n=1$) or a de Vaucouleurs one ($n=4$). 
To do that, we considered only the simulations belonging to these 
two classes, as we have seen in the previous section that
the true $n$ value can be retrieved for all the galaxies 
in the sample. The resulting number of de Vaucouleurs distributions 
is then 21 out of 41 (51\%). 

A more detailed classification was made leaving $n$ free to vary
among the integer values $n=1,2,3,4,5,6$. Using this approach,
the relative abundance of non--exponentials ($n\geq2$) is slightly 
higher if we choose the best $n$ value for each galaxy from the 
whole set of fits performed: 25/41 (61\%). In particular, four galaxies 
previously catalogued as exponentials are now fitted better by 
an $n=2$ distribution, whereas the other 10 objects with $n=1$ have 
their best fits confirmed.
As we discussed previously, for the galaxies placed 
above the dotted line in Fig.~\ref{fig_galps} we can reliably distinguish
between $n=1$ and $n\neq1$, whereas at fainter fluxes 
$n > 1$ distributions may be mistaken for exponentials. This 
``high--signal''
subsample, therefore, provides a particularly accurate estimate of the
fraction of likely ellipticals which, in this case, is even larger,
amounting to 81\% (21 out of 26).

As mentioned in Section 2, for $z\geq1$, the WFPC2 and NICMOS images 
cover the rest-frame UV and the optical spectral regions respectively. 
As it is well known that the galaxy morphology depends strongly 
on the wavelength (e.g. Kuchinski et al. \cite{kuchi} and references
therein), one may argue if this can have effects on our results. In this 
respect, we can envisage three cases. First, if a galaxy is a 
passively evolving elliptical, then its morphology does not depend 
on $\lambda$ (e.g. Kuchinski et al. \cite{kuchi}) and it would be 
classified as elliptical both in WFPC2 and in NICMOS images.
Second, if a galaxy is irregular, then it would be reliably
classified as such both in WFPC2 and in NICMOS images 
(e.g. HR10; Graham \& Dey \cite{grdey}; Dey et al. 
\cite{dey99}). Finally, there could be cases of elliptical
galaxies with a disk component having $n=1$ if observed in 
the optical (WFPC2) and $n\geq2$ if observed in the near-IR
(e.g. spheroidal galaxies with a disk component becoming more
prominent in the rest-frame UV). Our analysis does not allow to 
investigate if such latter cases are present in our sample because 
no NICMOS images are available for the 9 objects with $n=1$ 
observed with WFPC2. 

To summarize, we conclude that the galaxies that can be reliably 
classified as ellipticals amount to 50$\div$80\% of the total 
sample of 41 EROs. Although our sample is incomplete, we note that
our results are in good agreement with those of Stiavelli \& Treu 
\cite{st} based on a {\it complete} sample of NICMOS-selected EROs.

Such a high fraction of ellipticals strengthens the scenario proposed
by Daddi et al. \cite{daddi} who suggested that, because of their 
strong clustering, EROs are likely to be dominated by ellipticals 
rather than dusty starbursts. 

\section{Discussion \label{sect_disc}}

\subsection{Field and cluster objects}

Since our sample includes galaxies both from the field and from a cluster
environment, we can investigate the eventual differences between
these two subsamples. Following the conclusions of the authors, we will 
consider as cluster members the objects previously studied by 
Liu et al. \cite{liu} and S97, 
and assume the rest of the sample to be representative of the 
ERO field population; the two subsamples include 14 and 27 objects
respectively. 

The most remarkable difference between them is that most of the galaxies 
classified as irregular (5 out of 6) belong to the field population. 
On the other hand, the only irregular cluster object (n. 5 in Table 
\ref{table:oksample}) is neither diffuse nor characterized by two interacting 
components, but is rather a compact galaxy with an irregular core; also, 
its spectrum does not exhibit features typical of ongoing star formation (S97;
again, one example of an apparently ``old'' object that does not resemble 
local ellipticals). 
We conclude that, if some starburst 
galaxies are present among the EROs, they are not likely to be found in
clusters. Considering only the field population, the fraction of irregular
galaxies is only slightly increased (19\%) with respect to our previous
estimate.

For what concerns the fraction of non--exponential profiles, it is 
roughly the same for the two subsamples, again close to 80\% for the
high--signal objects.

In Table \ref{tab_stat} we report a summary of the sample statistics.
In the upper half of the table we consider the whole sample, 
divided into irregular galaxies, exponentials, and de Vaucouleurs.
In the lower half we consider only the high--signal subsample, for which
we distinguish between $n=1$ and $n>1$.

\begin{table}
\begin{flushleft}
\caption{ERO morphology: a summary}
\protect\label{tab_stat}
\begin{tabular}{l|llll}
\noalign{\smallskip}
\hline
\noalign{\smallskip}
\multicolumn{1}{c}{Whole sample} & \multicolumn{1}{c}{Total} & 
\multicolumn{1}{c}{Irr.} & \multicolumn{1}{c}{Exp. ($n=1$)} & 
\multicolumn{1}{c}{Ell. ($n=4$)} \\
\noalign{\smallskip}
\hline
\noalign{\smallskip}
All     & 41 & 6 (15\%) & 14 (34\%) & 21 (51\%) \\
Cluster & 14 & 1 (7\%)  & 5 (36\%) & 8  (57\%) \\
Field   & 27 & 5 (19\%) & 9 (33\%) & 13 (48\%)\\

\noalign{\smallskip}
\hline
\noalign{\smallskip}
\multicolumn{1}{c}{High signal} & \multicolumn{1}{c}{Total} & 
\multicolumn{1}{c}{Irr.} & \multicolumn{1}{c}{Exp. ($n=1$)} &         \multicolumn{1}{c}{Ell. ($n\geq 2$)} \\
\noalign{\smallskip}
\hline
\noalign{\smallskip}
All     & 26 & 1 (4\%) & 4 (15\%) & 21 (81 \%) \\
Cluster & 11 & 0 (0\%) & 2 (18\%) & 9  (82 \%) \\
Field   & 15 & 1 (7\%) & 2 (13\%) & 12 (80 \%) \\
\noalign{\smallskip}
\hline
%        &             &                &      \\
\end{tabular}
\end{flushleft}
 
%{\small Notes to Table~1.\\
%}
 
\end{table}
 
\subsection{Red exponential galaxies}

As we have seen, some of our galaxies appear to be compact exponentials
(see also Stiavelli \& Treu \cite{st}).
of course, these objects cannot be
classified as typical bright ellipticals if we use the local objects
as a reference; on the other hand, the regularity of their surface
brightness distributions tends to exclude the hypothesis of heavily
reddened objects. The existence of this subclass, therefore, implies that
the ERO population is apparently more composite than previously thought,
a conclusion also emerging from the work by Liu et al. \cite{liu}, 
Stiavelli \& Treu \cite{st} and Corbin et al. \cite{corbin}. 
The possibility that such objects are undergoing an
intermediate post--merging phase that eventually ends up in an elliptical
galaxy is in contrast with the simplest monolithic collapse model, in which
{\em all} ellipticals are formed at high redshift.
Its implications in the framework of the different
scenarios for galaxy formation certainly deserve further, more detailed
investigation.
At the same time, we cannot exclude that some ongoing star formation, 
suitably distributed throughout
the galaxy, might transform an elliptical--like bringhtness distribution
into one of the kind observed. Again, this hypothesis
could be tested by a better characterization of the stellar content of
these
exponential objects.

\subsection{Morphology and colors}

In Fig. \ref{fig_color} we plot the median $I-K$ colors, 
computed for the different morphological classes introduced: irregular
objects (corresponding to the $n=0$ point), exponential distributions, 
and the ones with $n > 1$. To include the galaxies with only $R-K$ 
available, we adopted a color $R-I=1.5$ to convert the $R-K$ color into
$I-K$. The adopted $R-I=1.5$ is an average value estimated for
elliptical galaxies using the synthetic spectral--energy distributions 
of Bruzual \& Charlot (1997) for a set of passively evolving 
models at $1<z<2$ (see the caption of Fig. 1). Since the $R-I$
color depends strongly on where the 4000~\AA~ break falls, the 
adopted $R-I=1.5$ has an associated uncertainty of $\pm 0.3$ mag.

The dotted line represents the average $I-K$ for $n \geq 2$ galaxies. 
Although each morphological bin is characterized by a significant 
scatter, there seems to be a tendence for the irregular objects 
to be characterized by the most extreme colors, with an average $I-K$
exceeding the mean color of likely ellipticals ($n>2$) by about 1 magnitude. 
Inverting the argument, we find that among the 6 reddest
EROs ($I - K > 5.5$) only one exhibits the typical morphology of an 
elliptical galaxy. 

The adopted $R-I=1.5$ is most likely appropriate for high redshift
ellipticals, but might be somewhat overestimated for possible starburst 
galaxies; thus, if we adopt a smaller $R-I$ for the irregular 
objects, the difference between them and the rest of the sample is 
furtherly increased. 
This result does not change also if we consider only the field subsample,
and it is broadly consistent with the recent findings that the EROs detected in
the submm show the reddest colors (Cimatti et al. \cite{cimatti};
Dey et al. \cite{dey99}; Smail et al. \cite{smail}; Gear et al.
\cite{gear}).
A possible connection -- to be confirmed by further, more detailed investigations --
is therefore suggested between morphology, submm emission, and optical/infrared colors 
(in particular $I-K$): high $z$ ellipticals and starburst might in fact exhibit a 
different behaviour with respect to each of the three parameters.

\begin{figure}
%\vspace{1.5cm}
\resizebox{\hsize}{!}{\includegraphics*[5mm,50mm][195mm,240mm]{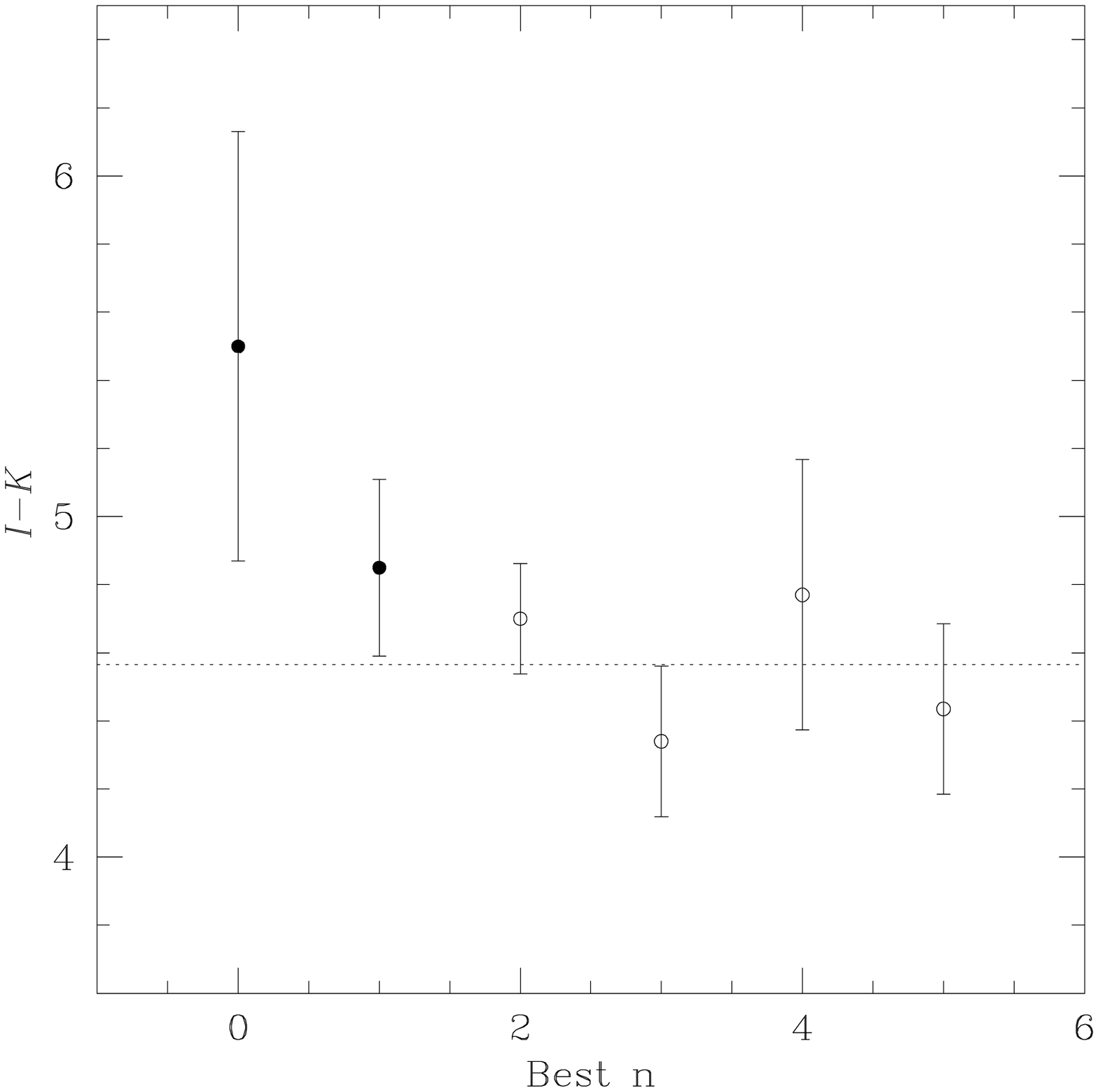}}
\caption{
Median $I-K$ colors for the different morphological classes; the $n=0$
point corresponds to the irregular objects of the sample. The horizontal 
dotted line is the average $I-K$ color for the $n\geq 2$ galaxies.
Irregular galaxies are characterized, on average, by the reddest colors.
}
\label{fig_color}
\end{figure}

\subsection{The distribution of $n$}

We can consider the subsample of galaxies whose best shape index $n$ is 
$\geq 2$ as a set of likely high--$z$ ellipticals. For these objects we can
compare the distribution observed for $n$ with the results
found for local samples of elliptical galaxies.
Caon et al. \cite{caon} determined a non--integer index $n$ for a sample of
local early--type galaxies in the Virgo cluster, so we 
have considered the galaxies from their sample with
$n_{best} \geq 1.5$, and rebinned them in the range 2--8. More recently,
the same kind of distribution was derived by Khosroshahi et al. \cite{khos}
for a sample of elliptical galaxies in the Coma cluster.

In the top panel of Fig.~\ref{fig_dist} 
we plot our distribution for the high--signal subsample
with no correction applied to the derived $n$ values (thick solid line).
The dotted line is the same distribution, corrected for the systematic 
effect described in Sect. \ref{sec_sim} (Eq. \ref{eq_cn}).
In the bottom panel we plot the distributions for the Virgo cluster 
and the Coma cluster.
Quite surprising, the two ``local'' histograms appear rather different, with 
the Virgo distribution extended up to large $n$ values, and clearly peaked at 
$n=2$, and the Coma distribution characterized by a peak and an upper 
cutoff at $n=4$. Due to this diversity a comparison with our data is rather 
difficult; we just note that our ``high--$z$'' distribution looks 
somewhat intermediate between the two, being quite flat between $n=2$ and
$n=5$, and confined to this range.
Without attempting any deeper comparison, we limit ourselves to 
consider this result as consistent with our claim that we are actually 
looking at a population of elliptical galaxies.
%We intreprete this result as 
%one supporting our claim that we are actually looking at a population of 
%true elliptical galaxies. 
 
\begin{figure}
%\vspace{1.5cm}
\resizebox{\hsize}{!}{\includegraphics*[5mm,50mm][195mm,240mm]{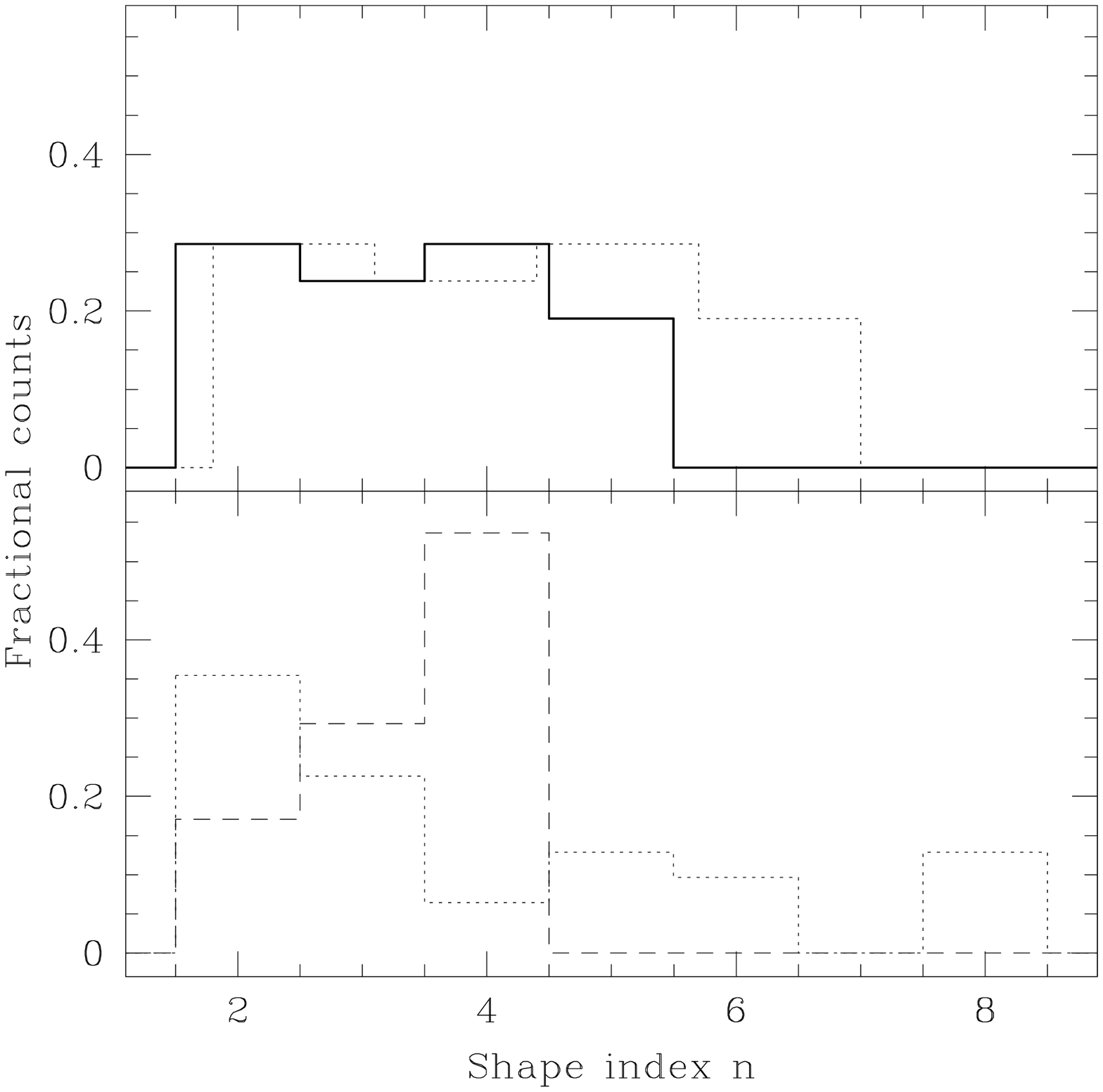}}
\caption{
Top panel: 
The distribution of the shape index $n$ derived from our sample, both
uncorrected (thick solid line) and corrected (dotted line) for the
systematic effect described in Sect. \ref{sec_sim}.
Bottom panel: the distributions derived by Caon et al. \cite{caon} (dotted
line), and by Khosroshahi et al. \cite{khos} (dashed line).
}
\label{fig_dist}
\end{figure}
 
\subsection{The Kormendy relation}

Scaling relations represent a powerful tool to investigate the evolution 
of galaxies at high redshift; in the case of elliptical galaxies,
the Kormendy Relation between effective surface brightness and radius
is relatively easy to build for a sample of ellipticals, when a detailed 
analysis of the brightness distributions and an accurate 
measure of the redshifts are available.
Previous studies of this kind (for example Fasano et al. \cite{fasano}, 
Ziegler et al. \cite{ziegler} -- Z99 hereafter, Roche et al. \cite{roche}) 
observed, as expected, an increase of the rest--frame surface brightness 
with redshift, but
the type of evolution implied (passive or partially active) is not yet well 
constrained by the models.
 
For 6 of our $n\geq 2$ compact galaxies a spectroscopic measure of the redshift 
has been published (in particular, 4 galaxies from S97 and 2 from Liu et al.
\cite{liu}). All these galaxies are likely to reside in a cluster
environment, and all of them are approximately at $z=1.3$. As a
consequence, they make up a particularly homogeneus set, well suited to 
pinpoint a particular time in the luminosity evolution of cluster 
ellipticals.  
Four more elliptical candidates (the ones in the HDFS) have photometric 
redshifts measured, but the two estimates available (Chen et al. \cite{chen}, 
B\'enitez et al.  \cite{benitez}) are quite discrepant and
they have been excluded from the following analysis.

We have used the 6 spectroscopic redshifts to derive the rest--frame
parameters of the relative galaxies, following the prescriptions
outlined in Z99; in particular, the observed F814W and H160W surface 
brightnesses have been corrected for the cosmological dimming, and 
transformed to the rest--frame $B$ band. 
The corrections (Pozzetti, private communication) are evaluated for 
different cosmologies and spectral templates, using the models described in 
Pozzetti et al. \cite{lucia}.
The observed luminosities have been corrected for the Galactic extinction
using the results by Schlegel et al. \cite{schlegel}.
In Fig. \ref{fig_korm} we plot the rest--frame Kormendy Relation for the 
6 selected galaxies, derived adopting $H_0 = 50$, $q_0 = 0.05$, and 
a single stellar population model with redshift of formation $z_f = 5$.
As a local reference, the solid line is the relation reported by Z99 for
the whole sample studied by J\o rgensen et al. \cite{joerg}:
\begin{equation}
\mu_e = 3.46 \, (+1.17, \, -0.73) R_e + 19.46 \, (+0.33, \, -0.55) \; \; .
\end{equation}

\begin{figure}
%\vspace{1.5cm}
\resizebox{\hsize}{!}{\includegraphics*[5mm,50mm][195mm,240mm]{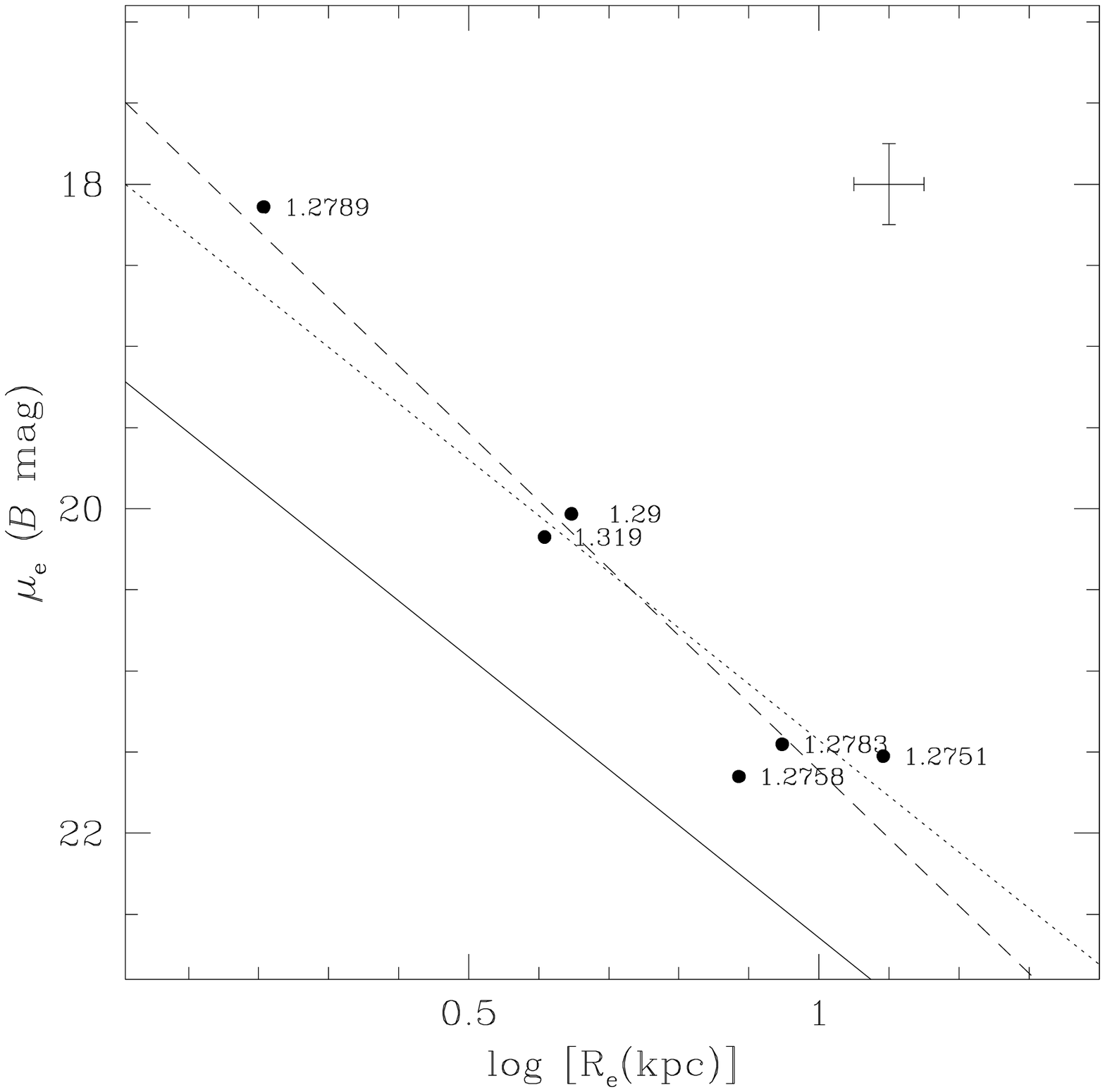}}
\caption{
The $B$--band, rest--frame Kormendy Relation for a subsample of ``cluster'' 
EROs at $z \sim 1.3$. The rest--frame parameters are evaluated 
assuming $H_0 = 50$ km s$^{-1}$ Mpc$^{-1}$ and $q_0 = 0.05$.
The solid line represents the local relation; 
the dotted line is the best fit to the data, with the slope fixed at the local
value; the dashed line is a fit with both parameters free. The typical 
uncertainty on the data points is shown in the upper right corner
of the plot.
}
\label{fig_korm}
\end{figure}

An upwards shift of the data points with respect to the local relation 
is evident: a line with the same slope fitted to the data (the dotted one 
in Fig. \ref{fig_korm}) yields a difference of 1.5 $\pm$ 0.4 mag. 
A value as low as 1.1 can be obtained by choosing a 
different local template (see Z99 for the details), or increasing the 
adopted value of $q_{\circ}$ up to 0.5.
The best fit with both parameters free (dashed line) is
\begin{equation}
\mu_e = (4.1 \pm 0.5) R_e + (17.45 \pm 0.15) \; \; ,
\end{equation}
consistent with the constant--slope hypothesis.
The possible slight steepening of this relation should be considered with 
caution, both because of the very few data points, and because of possible 
selection effects (for example, a set of galaxies 
selected in a narrow range of redshift and luminosity necessarily tends 
to exhibit a slope of 5).

A comparison with published results shows that
the measured shift in surface brightness is consistent with the
predictions of evolutionary models for elliptical galaxies (for example, 
with the Pure Luminosity Evolution models considered by Roche et al. 
\cite{roche}), as well as with the trends of luminosity
and surface brightness vs. $z$ observed at lower redshifts.
Z99, for example, find a difference of 0.8$\sim$0.9 $B$--mag for 
two clusters at $z=0.55$, whereas Schade et al. \cite{schade} estimate 
a luminosity evolution of about 1 mag for a sample of 
field galaxies at redshift between 0.75 and 1.

%(e.g.: Schade et al. \cite{schade}). 

\section{Summary and conclusions \label{sect_conc}}

We have implemented a code to analyze the surface brightness distribution
of faint galaxies, optimized to work on deep  HST images, and
we have tested its performances on a large set of simulated galaxies.
We have then used this code to fit model brightness distributions
to a sample of 41 EROs, imaged by HST in the optical or 
near--infrared wide--band filters, in order to identify a set of
high--$z$ elliptical candidates on a morphological basis.
The main results of this work can be summarized as follows.

a) We have determined the fraction of irregular objects
among the EROs of our sample, amounting to about 15\%. These galaxies
are the favourite candidates to be the hosts of dusty starbursts. 

b) Considering the whole sample, the galaxies characterized by a 
brightness distribution typical of local ellipticals are
at least 50\%. This estimate is based on a simple distinction between 
irregular/exponentials and de Vaucouleurs profiles; a more accurate 
analysis, allowing for different values of the shape index $n$, yields 
a higher fraction of ellipticals (70-80\%). 

c) We find that a rather small fraction of EROs (15 $\sim$ 30 \%) is made up 
of compact objects whose brightness distributions do not resemble the  
ones of local elliptical galaxies, but are better described by a pure
exponential law.

d) Our data suggest that irregular EROs are found predominantly in the 
field, and that they are characterized -- on average -- by the reddest
colors.
 
e) We have determined the rest--frame Kormendy relation for a subsample
of 6 cluster ellipticals, at redshift $\sim 1.3$. The relation turns out to be
brighter than the local one, at fixed size, by 1.4 mag in the $B$
band. 

In the near future, we plan to extend this work to larger,
complete samples of high redshift galaxies, in particular devoting our 
efforts to obtain a reliable picture of the whole population
of high--redshift ellipticals.

\begin{acknowledgements}

We are grateful to the referee, Dave Thompson, for his useful
comments, and to Lucia Pozzetti for the computations provided.
This work was partially funded by the C.N.A.A. grant n. 3/98 to G.M.

\end{acknowledgements}

\end{document}